\documentclass[11pt]{article}
\usepackage[utf8]{inputenc}
\usepackage[T1]{fontenc}
\usepackage{mathptmx}
\usepackage{microtype}
\usepackage[margin=1.1in]{geometry}
\usepackage{amsmath,amssymb}
\usepackage{graphicx}
\usepackage{float}
\usepackage{booktabs}
\usepackage{longtable}
\usepackage{calc}
\usepackage{array}
\usepackage{etoolbox}
\usepackage{xcolor}
\usepackage{listings}
\newcommand{\passthrough}[1]{#1}
\providecommand{\tightlist}{\setlength{\itemsep}{0pt}\setlength{\parskip}{0pt}}
\usepackage{fancyhdr}
\usepackage[hidelinks,pdftitle={State Without a Landlord: An Architecture Proposal for Peer-to-Peer Replication of Durable Workflow State},pdfauthor={Luca Maraschi; Matteo Collina},pdfsubject={Distributed systems; durable execution; peer-to-peer replication},pdfkeywords={durable execution, peer-to-peer, append-only logs, Hypercore, BitTorrent, content addressing, capability delegation, verifiable state}]{hyperref}
\usepackage{url}
\definecolor{codebg}{RGB}{247,247,249}
\definecolor{kw}{RGB}{0,60,150}
\definecolor{cm}{RGB}{90,110,90}
\definecolor{st}{RGB}{150,55,30}
\lstdefinelanguage{TypeScript}{
  keywords={type,const,let,var,function,async,await,return,if,else,for,of,in,new,class,implements,interface,extends,switch,case,break,throw,try,catch,private,public,readonly,import,from,export,null,true,false,number,string,boolean},
  keywordstyle=\color{kw}\bfseries,
  comment=[l]{//}, morecomment=[s]{/*}{*/}, commentstyle=\color{cm}\itshape,
  string=[b]', morestring=[b]", morestring=[b]`, stringstyle=\color{st}, sensitive=true }
\newcommand{\authorblock}{%
  \begin{tabular}[t]{c@{\hspace{3em}}c}
    {\large Luca Maraschi} & {\large Matteo Collina} \\[2pt]
    \normalsize Platformatic Inc. & \normalsize Platformatic Inc. \\
    \normalsize\ttfamily luca@platformatic.dev & \normalsize\ttfamily matteo@platformatic.dev
  \end{tabular}}
\makeatletter
\def\maketitlepage{%
  \begin{center}
  {\LARGE\bfseries State Without a Landlord:\\[2pt]
   An Architecture Proposal for Peer-to-Peer\\[2pt]
   Replication of Durable Workflow State\par}
  \vspace{14pt}
  \authorblock\par
  \vspace{8pt}
  {\normalsize September 2026\par}
  \end{center}\vspace{6pt}}
\makeatother
\begin{document}
\maketitlepage
\thispagestyle{plain}
\begin{abstract}\noindent
Durable-execution frameworks commonly journal execution steps of a long-running function and reconstruct state by deterministic replay; the journal is therefore the workflow's authoritative state, and in current deployments it typically resides with whichever provider hosts the run. This paper examines what changes if the journal is instead an authenticated, append-only log replicated among the parties to the workflow. We propose an architecture in which runs are chains of single-writer epoch cores; executor succession is governed by quorum-finalized closure certificates; durability, timeout, and checkpoint decisions are carried by explicit certificates rather than implicit trust; and large payloads travel a separate content-addressed distribution plane. A threat model separates what signatures establish (authorship and order) from what they do not (truth of recorded effects). We state the design as a set of typed artifacts and invariants, identify which claims are established by existing systems and which are proposals requiring validation, and define the prototype experiments, including specific adversarial cases, that would validate or falsify the proposal's central mechanisms.
\end{abstract}

{\small\tableofcontents}

\hypertarget{the-problem-durable-execution-re-centralizes-state}{%
\section{The Problem: Durable Execution Re-Centralizes
State}\label{the-problem-durable-execution-re-centralizes-state}}

\hypertarget{what-durable-execution-actually-is}{%
\subsection{What durable execution actually
is}\label{what-durable-execution-actually-is}}

Strip away the ergonomics and a durable workflow engine is an
event-sourced interpreter. In Workflow SDK (Vercel's durable-execution
framework, formerly the Workflow Development Kit), a function marked
\passthrough{\lstinline!"use workflow"!} is compiled into a
deterministic orchestrator; functions marked
\passthrough{\lstinline!"use step"!} are the effectful leaves. Each
step's input and output is recorded in an event log. On resumption,
after a crash, a deploy, or a \passthrough{\lstinline!sleep("7d")!}, the
orchestrator re-executes from the top, and each
\passthrough{\lstinline!await someStep(...)!} returns the cached result
from the log instead of re-running. Nondeterminism sources
(\passthrough{\lstinline!Date.now()!},
\passthrough{\lstinline!Math.random()!}) are intercepted and pinned
during replay so every replay makes the same decisions.

Three properties fall out of this design, and everything that follows
rests on them:

\begin{enumerate}
\def\labelenumi{\arabic{enumi}.}
\tightlist
\item
  \textbf{The journal is the workflow.} Code plus journal fully
  determines state. The running process is disposable; the log is not.
\item
  \textbf{The journal is append-only and (per run) single-writer.}
  Exactly one executor advances a given run at a time; it only ever
  appends step records.
\item
  \textbf{Replay is deterministic.} Any party holding the code version
  and the journal can reconstruct the workflow's state: it can not
  merely observe the run but \emph{become} its executor.
\end{enumerate}

These are the very properties that make the architecture durable. They
are also, as we will show, the same properties that make the journal a
natively replicable P2P object.

\hypertarget{where-the-journal-lives-today}{%
\subsection{Where the journal lives
today}\label{where-the-journal-lives-today}}

In every production deployment of a durable execution system, the
journal lives in a centrally owned store: the vendor's managed
infrastructure (Redis-backed streams and distributed queues in Vercel's
hosted World), a cloud primitive (Durable Objects storage in Cloudflare
Workflows), or a self-hosted database (Temporal's history service,
Platformatic's Kubernetes World). The \passthrough{\lstinline!World!}
abstraction in Workflow SDK is explicit about this: a World defines how
execution, orchestration, and persistence are handled, and every
existing World resolves persistence to a single administrative domain.

This has four consequences, in ascending order of severity.

\textbf{Lock-in of state, not code.} Durable execution vendors correctly
advertise ``no lock-in'' at the code level: the same workflow runs on
any World. But a \emph{run in flight} is not portable. Migrating
providers mid-run means either draining every workflow to completion
(impossible for workflows that sleep for months) or exporting and
re-importing journals across incompatible stores. Code portability
without state portability is portability for greenfield only.

\textbf{Availability coupling.} A workflow that could in principle
resume anywhere can in practice resume only where its journal is
reachable. The journal store's availability upper-bounds the workflow's
availability. Note the perversity: the whole point of durable execution
is surviving infrastructure failure, yet the design concentrates the one
irreplaceable artifact in one infrastructure.

\textbf{Unverifiable history.} The journal is plain rows in an
operator's database. Nothing binds the recorded history
cryptographically to what actually executed. The operator (or anyone who
compromises the store) can rewrite step results, and replay will
faithfully reconstruct the forged state. For a single-tenant app this is
a theoretical concern. For workflows that encode obligations between
parties (payments, approvals, agent-to-agent contracts), it means the
system's memory is only as trustworthy as one party's database.

\textbf{Multi-party workflows require a landlord.} Consider a workflow
spanning two organizations, or two autonomous agents operated by
different principals: an approval gate where org A requests and org B
approves; a saga whose compensating actions cross a company boundary; an
agent negotiation with staged commitments. Today, someone must host the
run, and everyone else must trust the host's record of it. The workflow
has a landlord, and the landlord's ledger is the truth. This is
precisely the structure that centralized package registries impose on
software distribution, and it has the same failure modes: the landlord
is a single point of censorship, tampering, outage, and rent extraction.

A note on the word, since much in this paper depends on it.
\emph{Landlord} is a technical term here, not an accusation: a party
whose custody of state is simultaneously \textbf{exclusive} (no other
party holds an authoritative copy), \textbf{required for liveness} (the
workflow cannot advance without the custodian's cooperation), and
\textbf{non-substitutable} (replacing the custodian requires the
custodian's help). The definition names a structural position, not
conduct: most operators occupying it are competent, honest, and cheap,
and that is why the position's failure modes stay invisible until the
moment they bind. The argument of this paper is never that hosted
operators misbehave; it is that a design where only trust prevents the
failure modes is strictly weaker than one where the position does not
exist. And the claim is constructive, not ascetic: §10 argues that
everything valuable such operators provide (speed, wake reliability,
observability, indexes) survives as competitive services on the open
substrate once the position itself is dissolved.

\textbf{What this paper is, and is not.} This document is an
architecture proposal: it argues that a design point exists and is worth
building, and it specifies that design precisely enough to be criticized
and prototyped. It is not a protocol specification, a security proof, or
an experience report. No implementation of this architecture yet exists.
To keep that discipline at the sentence level, the paper classifies its
claims throughout: \textbf{{[}E{]}stablished} marks properties that
follow from the documented behavior of existing systems (Hypercore's
signed Merkle logs, BitTorrent v2's content addressing, Workflow SDK's
replay semantics); \textbf{{[}P{]}roposed} marks mechanisms this paper
designs, whose stated properties are design intent pending validation;
\textbf{{[}O{]}pen} marks problems where even the design is unsettled.
Where a stronger word might tempt, the paper uses the accurate one:
proposed, with the validation path named in §13.

\hypertarget{the-agentic-forcing-function}{%
\subsection{The agentic forcing
function}\label{the-agentic-forcing-function}}

Human-operated systems tolerate landlords because humans move slowly and
litigate disputes out of band. Agentic systems do not. When workflows
are initiated, executed, and consumed by autonomous agents at machine
speed (often agents belonging to different principals, composing each
other's services dynamically) three assumptions behind centralized state
ownership break simultaneously: that the set of parties is known in
advance (so a host can be pre-agreed), that disputes are rare (so an
unverifiable log is acceptable), and that runs are short (so lock-in of
in-flight state is tolerable). Agent-to-agent workflows are long-lived,
adversarial-by-default, and dynamically composed. They need workflow
state that is \emph{jointly held, independently verifiable, and
executor-portable}. That is a P2P replication problem, and it has been
solved (twice) by communities that were not thinking about workflows at
all.

\begin{center}\rule{0.5\linewidth}{0.5pt}\end{center}

\hypertarget{related-work}{%
\subsection{Related work}\label{related-work}}

The mechanisms proposed here descend from named lineages, and the debts
should be explicit. Fencing by epoch (§8.1) is the fencing-token pattern
of coordination services (Chubby's lock sequencers and the token
discipline popularized by Kleppmann) materialized as log topology
instead of storage-side checks. The quorum-intersection rule (§8.4) is
Gifford's weighted-voting insight applied to the intersection of a
durability quorum with a reconfiguration quorum, and epoch transitions
themselves are a reconfiguration problem with a long consensus
literature (PBFT's view changes, Raft's joint consensus) that §13.7 must
engage seriously the moment indexers stop being crash-only. Capability
delegation follows the object-capability tradition (Dennis--Van Horn
through UCAN). The journal-as-replicated-log framing owes to the
event-sourcing and log-centric-systems literature, and the closest P2P
ancestors are Secure Scuttlebutt's signed single-writer feeds, OrbitDB's
CRDT stores over IPFS, and the Hypercore ecosystem this design builds on
directly, none of which, to our knowledge, couples an authenticated log
to a deterministic-replay execution contract with fenced writer
succession, which is the specific composition proposed here.

\hypertarget{the-isomorphism-workflow-journals-are-p2p-logs}{%
\section{The Isomorphism: Workflow Journals Are P2P
Logs}\label{the-isomorphism-workflow-journals-are-p2p-logs}}

\hypertarget{hypercore-briefly}{%
\subsection{Hypercore, briefly}\label{hypercore-briefly}}

Hypercore (the foundation of the Pear runtime's stack) is a secure,
distributed append-only log. A core is identified by a public key; only
the holder of the corresponding private key can append; every block is
signed, and the log's integrity is protected by a Merkle tree, so any
peer can sparsely replicate any subset of blocks and verify them against
the root without trusting the peer it fetched from. Replication is live:
peers subscribed to a core receive appends as they happen. Fork
detection is built in: a writer that rewrites history produces
cryptographic evidence of its own misbehavior, because two signed
messages for the same sequence number constitute a proof of
equivocation.

On top of Hypercore the stack provides Hyperbee (an append-only B-tree
over a core, for sorted key-value data), Hyperdrive (a filesystem
abstraction), Hyperswarm (topic-based peer discovery over a Kademlia
DHT, with NAT holepunching), Corestore (management of many related
cores), and (critically for us) Autobase, a multi-writer layer that
linearizes many single-writer cores into one deterministic view via a
causal DAG.

\hypertarget{the-mapping}{%
\subsection{The mapping}\label{the-mapping}}

Set the two systems side by side and the correspondence is not
analogical but structural:

\begin{longtable}[]{@{}
  >{\raggedright\arraybackslash}p{(\columnwidth - 4\tabcolsep) * \real{0.3333}}
  >{\raggedright\arraybackslash}p{(\columnwidth - 4\tabcolsep) * \real{0.3333}}
  >{\raggedright\arraybackslash}p{(\columnwidth - 4\tabcolsep) * \real{0.3333}}@{}}
\caption{The isomorphism, feature by feature (§2.2).}\tabularnewline
\toprule\noalign{}
\begin{minipage}[b]{\linewidth}\raggedright
Durable execution concept
\end{minipage} & \begin{minipage}[b]{\linewidth}\raggedright
P2P primitive
\end{minipage} & \begin{minipage}[b]{\linewidth}\raggedright
Notes
\end{minipage} \\
\midrule\noalign{}
\endfirsthead
\toprule\noalign{}
\begin{minipage}[b]{\linewidth}\raggedright
Durable execution concept
\end{minipage} & \begin{minipage}[b]{\linewidth}\raggedright
P2P primitive
\end{minipage} & \begin{minipage}[b]{\linewidth}\raggedright
Notes
\end{minipage} \\
\midrule\noalign{}
\endhead
\bottomrule\noalign{}
\endlastfoot
Run journal (event log of step records) & Hypercore & Both append-only;
both single-writer; both totally ordered per log \\
Executor identity & Core keypair & The append key \emph{is} the right to
advance the run \\
Journal integrity & signed Merkle-tree states (signed lengths)
authenticating every block & Upgrade: today's journals have no integrity
story at all \\
State reconstruction via deterministic replay & Deterministic apply over
a verified log & Same discipline; Hypercore adds that the input log is
tamper-evident \\
Resumption after crash & Sparse/live replication + replay & Any peer
holding the log can resume, not just the original host \\
Journal store & Swarm of interested peers & Availability is a function
of interest, not of one operator's uptime \\
Executor rewrites history (failure mode) & Fork proof (equivocation is
detectable) & Centralized stores make this silent; Hypercore makes it
evidence \\
Multi-writer coordination (approvals, hooks, agents) & Autobase causal
DAG + deterministic apply & See §6 \\
Large step payloads / artifacts & BitTorrent v2 objects by infohash &
See §2.3, §4.2 \\
Discovery (``where is run X?'') & DHT topic = hash of run/core key &
Hyperswarm or BitTorrent DHT \\
\end{longtable}

The single-writer constraint matters more than it looks: it is where
naive decentralization proposals usually die. Multi-master replication
of workflow state would be a consistency nightmare, but durable
execution \emph{already forbids it}. Exactly one executor advances a
run; everyone else is a reader until an explicit handoff. Hypercore's
one-keypair-one-writer model is not a limitation to engineer around; it
is the same invariant the workflow engine already enforces, now enforced
cryptographically instead of by a lease row in somebody's database.

Similarly, deterministic replay (the discipline durable execution
imposes on orchestrator code) is the same discipline Autobase imposes on
its \passthrough{\lstinline!apply!} function: given the same linearized
inputs, every peer must compute the same view. Developers who have
internalized ``no \passthrough{\lstinline!Date.now()!} in workflow
code'' have already internalized the contract a P2P deterministic-apply
layer requires. The two communities converged on the same constraint
because it is the price of ``state = fold(log)'', and both were buying
that property.

\hypertarget{why-bittorrent-and-the-pear-stack}{%
\subsection{\texorpdfstring{Why BitTorrent \emph{and} the Pear
stack}{Why BitTorrent and the Pear stack}}\label{why-bittorrent-and-the-pear-stack}}

The two lineages are not competitors here; they answer different
questions, and the split rests on a principle this paper treats as
axiomatic: \textbf{transport and trust are separate concerns; likewise,
live coordination and bulk distribution are separate concerns.}
Conflating them (making the party that moves the bytes also the arbiter
of what the bytes mean) is precisely how centralized registries and
hosted workflow stores become landlords, so the architecture keeps the
planes apart from the first primitive.

Hypercore is optimized for small, hot, mutable-by-append data with live
subscribers -- just the shape of a journal receiving step records every
few seconds and readers who want them immediately. BitTorrent v2 is
optimized for large, cold, immutable data with massive fan-out, per-file
Merkle piece trees, content addressing by infohash, a battle-tested DHT,
and two decades of swarm behavior under adversarial conditions. A
workflow journal should not carry a 2 GB model checkpoint or a container
image inline; it should carry a 32-byte infohash, and the payload should
travel through the swarm layer. The journal stays small enough to
replicate to every interested peer in milliseconds; the artifacts flow
through infrastructure designed for that job.

This yields a clean two-plane architecture: a \textbf{coordination
plane} (Hypercore/Autobase journals, small, live, signed, totally
ordered) and a \textbf{distribution plane} (BitTorrent v2 swarms, large,
immutable, content-addressed), joined by hash references. The same split
generalizes beyond execution state (§11 sketches how) but this paper
needs only the workflow instance of it.

\begin{center}\rule{0.5\linewidth}{0.5pt}\end{center}

\hypertarget{threat-and-failure-model}{%
\section{Threat and Failure Model}\label{threat-and-failure-model}}

Every guarantee claimed later in this paper is relative to the
assumptions stated here; a mechanism that is sound under crash faults
and unsound under Byzantine faults is described as such. The model also
fixes an important scoping fact at the outset: \textbf{signatures and
Merkle proofs establish authorship and log integrity, never truth.} A
malicious executor can append a perfectly valid, non-forking
\passthrough{\lstinline!step.end!} recording an outcome that never
happened, and every peer will faithfully verify and preserve the lie.
Cryptographic replication authenticates \emph{who said what, in what
order}; whether the recorded effects occurred is a separate problem,
addressed only partially here (idempotent effect discipline, §8.6) and
otherwise deferred to external attestation (effect-target receipts, TEEs,
§13).

\textbf{Adversary classes.} The architecture targets three regimes, in
increasing order of difficulty, and each mechanism names the regime it
is designed for:

\textbf{T1: Crash-only (the design's primary target).} Executors,
participants, and indexers may crash, restart, lose disks, and
partition, but follow the protocol while running. Under T1 the
architecture aims for: no lost or corrupted run state given surviving
replicas; resumability by any authorized keyed peer; and detection of
accidental divergence (version skew, nondeterminism) at resumption seams
via transcript commitments (§7).

\textbf{T2: Rational/misbehaving participants.} Parties may equivocate,
lie about outcomes, claim falsely, or shirk paid obligations, but do not
control the network. Under T2 the architecture aims for \emph{evidence},
not prevention: equivocation yields fork proofs; custody shirking yields
evidenced breach claims (§12); recorded outcomes remain unverified
absent external attestation, and the paper says as much wherever it
matters. Executor eviction under T2 is a \emph{policy decision under
stated synchrony assumptions} (§8.3), not a proven fault determination.

\textbf{T3: Byzantine coordination and network adversaries.} Malicious
indexer majorities in a coordination Autobase, Sybil floods of a swarm,
global passive traffic analysis, and partition-controlling network
adversaries are \textbf{out of scope for this proposal's guarantees}.
The design degrades in stated ways under T3 (a corrupt indexer quorum
can finalize a bad handoff; a partition can cause false eviction whose
blast radius §8.3 bounds but does not eliminate), and hardening against
T3 (BFT indexer sets, stake, external anchoring) is future work (§13).

The regimes, their target guarantees, and their exclusions in one view:

\begin{longtable}[]{@{}
  >{\raggedright\arraybackslash}p{(\columnwidth - 4\tabcolsep) * \real{0.3333}}
  >{\raggedright\arraybackslash}p{(\columnwidth - 4\tabcolsep) * \real{0.3333}}
  >{\raggedright\arraybackslash}p{(\columnwidth - 4\tabcolsep) * \real{0.3333}}@{}}
\toprule\noalign{}
\begin{minipage}[b]{\linewidth}\raggedright
Regime
\end{minipage} & \begin{minipage}[b]{\linewidth}\raggedright
Target guarantees
\end{minipage} & \begin{minipage}[b]{\linewidth}\raggedright
Explicitly out of scope
\end{minipage} \\
\midrule\noalign{}
\endhead
\bottomrule\noalign{}
\endlastfoot
T1 crash-only & No lost or corrupted state given surviving replicas;
resume by any authorized keyed peer; divergence detected at resumption
seams (§7) & (design target) \\
T2 rational / misbehaving & Evidence, not prevention: fork proofs,
closure/timeout/ack/checkpoint certificates, audit trails; false
eviction bounded and recoverable (§8.3) & Truth of recorded effects
(§3); prevention of lying executors \\
T3 Byzantine coordination / network & Stated degradations only & Corrupt
indexer or checkpoint-quorum majorities; Sybil floods;
partition-controlling and globally observing network adversaries \\
\end{longtable}

\textbf{Failure assumptions made explicit.} \emph{Partial synchrony:}
liveness mechanisms (challenge windows, eviction, wake escalation)
assume message delays and clock skew are bounded by known constants
\emph{most of the time}; during violations, safety must not depend on
timing; the design goal is that timing violations cause delay or
recoverable duplication, never state corruption. \emph{Key compromise:}
a stolen executor key is equivalent to a malicious executor until
rotated; a stolen participant KEK exposes payload plaintext irrevocably
for data already replicated (§9.3 states the erasure consequences
without varnish). \emph{External effects:} target services are assumed
to either honor idempotency keys, be safely re-queryable, or be
explicitly marked as neither, in which case the architecture surfaces
uncertainty rather than hiding it (§8.6). \emph{Availability:}
replication-by-interest (P5) is a heuristic that concentrates copies
where stakes are, not a durability guarantee; contracted custody (§12)
exists precisely because the heuristic has gaps. \emph{Code trust:} the
runtime provides no isolation; an executing peer runs the
manifest-pinned bundle with the host's full authority, so resuming a run
means trusting its code (§9.10 states the gap and the proposed
step-ownership discipline).

\begin{center}\rule{0.5\linewidth}{0.5pt}\end{center}

\hypertarget{architecture-the-p2p-world}{%
\section{Architecture: The P2P World}\label{architecture-the-p2p-world}}

Workflow SDK's \passthrough{\lstinline!World!} interface is the
sanctioned seam: a World supplies storage, queuing, and authentication,
and the same workflow code runs unchanged against any World (Local in
development, Vercel's hosted World in production, Platformatic's
Kubernetes World self-hosted). We define a fourth: the \textbf{P2P
World}, in which those three responsibilities resolve to swarm
primitives instead of an administrative domain.

\hypertarget{layer-stack}{%
\subsection{Layer stack}\label{layer-stack}}

\begin{table}[H]\centering\small
\begin{tabular}{@{}ll@{}}
\toprule
\textbf{Execution layer} & Workflow SDK (\texttt{"use workflow"}/\texttt{"use step"}), unchanged \\
\textbf{Journal layer} & A chain of single-writer epoch cores per run (\S8.1): signed, append-only \\
\textbf{Coordination layer} & Autobase over participant cores, used within its documented guarantees (\S6) \\
\textbf{Distribution plane} & BitTorrent v2 swarms for large payloads; journal carries infohashes \\
\textbf{Discovery layer} & Hyperswarm / DHT: topic = run key; BitTorrent DHT: infohash lookup \\
\textbf{Trust layer} & Keypairs, delegation capabilities, fork proofs; optional attestation log (\S11) \\
\bottomrule
\end{tabular}
\caption{The P2P World layer stack (\S4.1).}
\end{table}

\begin{figure}[H]\centering\includegraphics[width=0.98\linewidth]{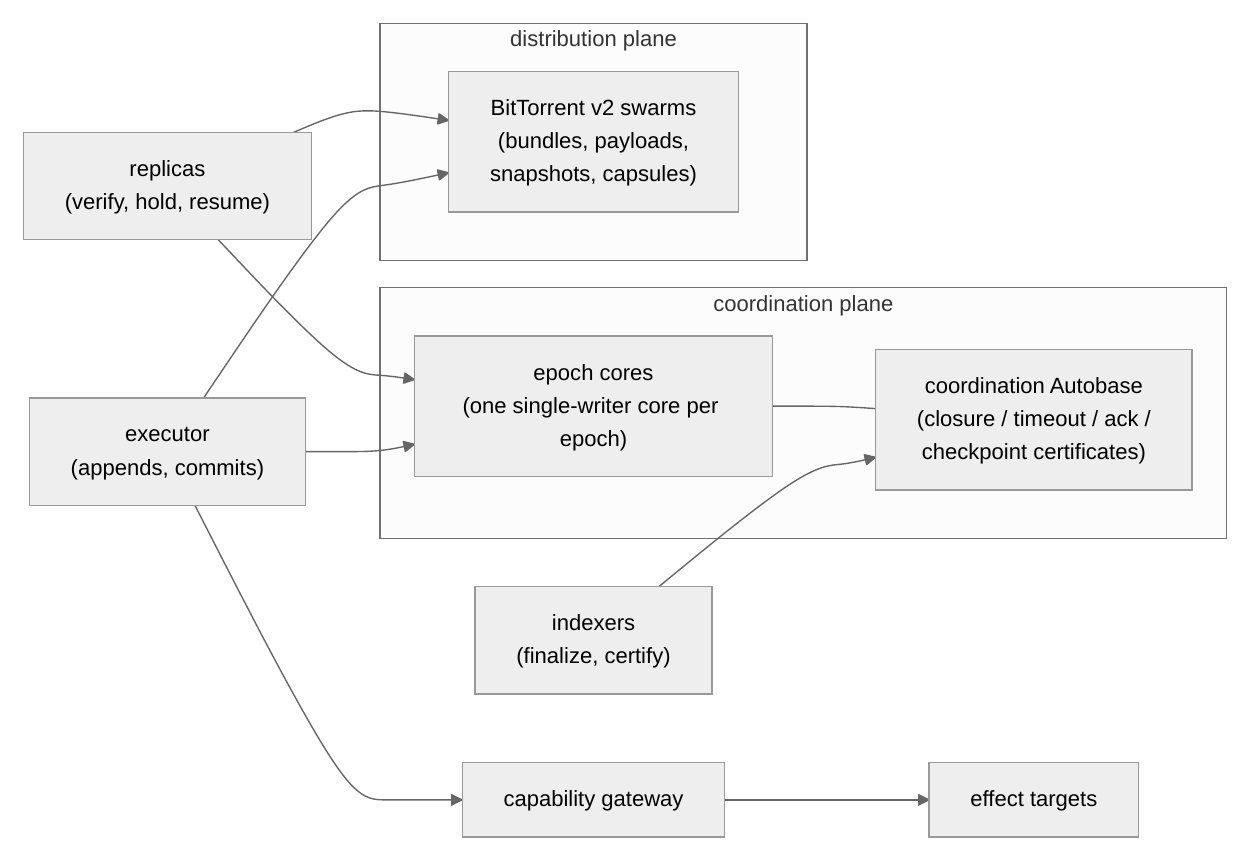}\caption{System overview: planes, actors, and the certificates that connect them (\S4.1).}\end{figure}

\hypertarget{anatomy-of-a-step}{%
\subsection{Anatomy of a step}\label{anatomy-of-a-step}}

Seen from a single step's perspective, the two-plane split of §2.3
works as follows: the executor appends a small signed fact whose bulk
payload rides the distribution plane, and both propagate to whoever
holds a stake:

\begin{figure}[H]\centering\includegraphics[width=0.95\linewidth]{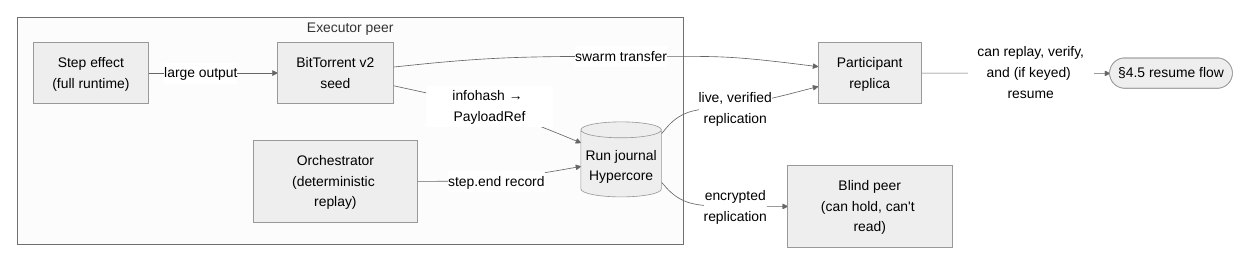}\caption{The two-plane anatomy of a single step (\S4.2).}\end{figure}

\hypertarget{data-model}{%
\subsection{Data model}\label{data-model}}

\textbf{The identity model.} Four distinct objects are easy to blur
together, and the design separates them explicitly \textbf{{[}P{]}}: (1)
the \textbf{run identity} \passthrough{\lstinline!runId!} is the genesis
epoch core's public (verification) key, the stable name of the chain,
not of any writer; (2) each \textbf{epoch core key} (§8.1) verifies
appends within one epoch, and its secret half is the current executor's
append authority for that epoch only; (3) the \textbf{executor identity}
is a principal keypair, long-lived across runs, named in the manifest
and bound to epoch keys by signature; (4) \textbf{participant
identities} are principal keypairs entitled to replicate, decrypt (per
policy), and co-sign. The manifest binds (1), (3), and (4); each
\passthrough{\lstinline!epoch.open!} binds a new (2) to (3). No key
plays two roles. One further distinction, forced by the fork machinery
of §8: \textbf{epoch identity}
\passthrough{\lstinline!(runId, epochNumber, coreKey)!} is stable across
every branch the epoch key ever signed, while \textbf{branch identity}
\passthrough{\lstinline!(forkId, treeRoot)!} names one fork lineage, and
every safety decision (unique succession, certificate lookup) keys on
the former, never the latter, so that no artifact of \emph{which branch
a message names} can hide a certificate or admit a second successor.

A run is born as a \passthrough{\lstinline!run.manifest!}, the unit of
the coordination plane, binding together everything a stranger needs to
verify and (given authorization) execute the run:

\begin{lstlisting}
type RunManifest = {
  runId: string;                  // = journalCore public key, hex
  workflow: {
    name: string;
    codeVersion: {
      infohash: Infohash;         // BitTorrent v2 hash of the built workflow bundle
      entrypoint: string;
    };
    worldContract: "p2p-world/0"; // semantics version of this spec itself
  };
  executor: {
    identity: PublicKey;          // GENESIS executor's principal identity;
                                  // succession is epoch.open (S8.1), never
                                  // manifest mutation - this field is immutable
    delegationChain?: Capability[]; // if executing on behalf of a principal
  };
  participants: PublicKey[];      // parties entitled to replicate & verify
  policy: {
    payloadEncryption: "none" | "participants" | "per-subject";
    retention: RetentionPolicy;     // see S9.3
    ack: "local" | { replicas: number } | "survivable";  // S8.4, S9.6
    liveness: "best-effort" | "mutual-stake" | "contracted"; // see S9.2
  coordination: { baseKey: CoreKey;          // run Autobase bootstrap (S6)
    ackThreshold?: number;        // default max(ceil((N+1)/2), N-q+1);
                                  // T2 manifests: > (N+f)/2 for assumed f (S8.4)
                  indexers: PublicKey[];      // finality quorum members
                  quorum: number };           // majority threshold
    failover: PublicKey[];          // pre-delegated standby executors, see S8
    livenessTiming?: {              // floors are normative (S8.3); omitted =
      grace: Duration;              //   the liveness/standard-0 profile
      challengeWindow: Duration; cooldown: Duration };
    readiness?: "declaration" | "challenge"; // successor claim strength (S8.3);
                                    // default "declaration" (sufficient under T1)
    attestation: {
      commitEveryN: number;       // executor attests state hash each N entries
      checkpointQuorum: number;     // co-signatures licensing pruning. NOT a
                                    // free dial: T1 tolerates any value; a
                                    // manifest claiming T2 durability with up
                                    // to f malicious checkpoint signers MUST
                                    // set checkpointQuorum > f (one honest
                                    // signer performed the S7.2 duties);
                                    // default: the coordination finality
                                    // quorum. Below that bound, a lying
                                    // checkpoint is T2 behavior, not T3.
    };
  };
  createdAt: HLC;                 // hybrid logical clock timestamp
  signature: Signature;           // by the initiating principal
};
\end{lstlisting}

Journal entries are step records, each self-describing and each
referencing bulk data out-of-plane:

\begin{lstlisting}
// Every entry additionally carries { hlc: HLC } - the executor's hybrid
// logical timestamp at append - omitted below only for brevity. [P]
type JournalEntry =
  | { t: "run.start";    manifest: RunManifest }
  | { t: "step.begin";   seq: number; step: string; inputRef: PayloadRef;
      attempt: number;
      callSite: bytes; callIndex: number; argsHash: Hash }  // intent trace (S7.1)
  | { t: "step.end";     seq: number; step: string; outputRef: PayloadRef;
      attempt: number; hlc: HLC }
  | { t: "step.error";   seq: number; step: string; error: ErrorRef;
      attempt: number; willRetry: boolean }
  | { t: "sleep.begin";  seq: number; wakeAfter: HLC;
      callSite: bytes; callIndex: number; argsHash: Hash }  // intent trace (S7.1)
  | { t: "hook.await";   seq: number; hookId: string; grantedTo: PublicKey[];
      callSite: bytes; callIndex: number; argsHash: Hash;     // intent trace (S7.1)
      base: { key: CoreKey; length: number; viewRoot: Hash } } // Autobase pin (A.4)
  | { t: "hook.receive"; seq: number; hookId: string; payloadRef: PayloadRef;
      from: PublicKey; signature: Signature }
  | { t: "manifest.ack";  chainHead: Hash }                 // amendments enter the
                                                            // fold here (S10.3, A.3)
  | { t: "child.spawn";  coreKey: CoreKey;                  // run fission (S9.9)
      callSite: bytes; callIndex: number; argsHash: Hash }  // intent trace (S7.1)
  | { t: "child.join";   coreKey: CoreKey; checkpointHash: Hash }
  | { t: "transcript.commit"; seq: number; stateHash: Hash;   // S7.1
      nextIntent: NextIntent; engine: Infohash }
  | { t: "checkpoint"; cert: CheckpointCertificate }         // S7.2 - the
                                  // co-signed binding of prefix, snapshot,
                                  // capsule, state, and next intent
  | { t: "run.end";      result: PayloadRef | ErrorRef;
      callSite: bytes; callIndex: number; argsHash: Hash }; // intent trace, kind "end":
                                  // callSite = the completion point's bundler id;
                                  // argsHash = hash of the canonical result commitment

type NextIntent = { kind: "step" | "sleep" | "hook" | "child" | "end";
                    callSite: bytes; callIndex: number; argsHash: Hash };
// seq       = position in the canonical run stream (S8.1) - journal order.
// callIndex = ordinal in the orchestrator's deterministic call sequence.
// The two are related but not equal: retries advance seq, not callIndex.

type PayloadRef =
  | { kind: "inline"; keyId: KeyId;              // small values, <= ~4 KB -
      nonce: Uint8Array; ct: Uint8Array }        // enveloped like blobs (S9.1)
  | { kind: "blob";   infohash: Infohash;        // BitTorrent v2 object
      size: number; encryption?: EncryptionHeader };
\end{lstlisting}

A note on \passthrough{\lstinline!Infohash!} semantics \textbf{{[}E{]}}:
a BitTorrent v2 infohash commits to the \emph{bencoded info dictionary}
of a metainfo, not directly to a raw byte string.
\passthrough{\lstinline!PayloadRef.blob!} therefore denotes the infohash
of a canonical single-file v2 metainfo wrapping the payload (fixed piece
size, no optional fields) so that payload bytes map to exactly one
infohash; that canonical wrapping is part of the
\passthrough{\lstinline!worldContract!} and belongs in the conformance
corpus (§13).

Every entry is appended to the current epoch's core by the executor. The
core's Merkle tree gives any replica an inclusion proof for any entry;
the signing key ties every append to the epoch whose
\passthrough{\lstinline!epoch.open!} bound it to an executor identity
(§8.1) -- for epoch 0, the identity the manifest names.

\hypertarget{the-world-interface-a-hypothesis-sketched}{%
\subsection{The World interface: a hypothesis,
sketched}\label{the-world-interface-a-hypothesis-sketched}}

The headline promise (workflow code unchanged, only a new World) is a
\textbf{hypothesis, not a result {[}O{]}}, and the gap should be sized
frankly. The real \passthrough{\lstinline!@workflow/world!} contract is
substantially wider than the four methods sketched below: it spans
queue, stream, and storage surfaces and includes protocol-version
negotiation, event creation, materialized run/step entities, hook
deduplication, stream lifecycle, deployment affinity, encryption-key
lookup, and atomicity requirements on specific operations. Some of those
surfaces map naturally onto the primitives here (streams onto cores,
§10.1; storage onto the journal; hook dedup onto
\passthrough{\lstinline!hookId!} in the fold); others (atomic
multi-entity updates, affinity) need design work this paper has not
done. And an adapter is not the whole cost: the architecture also
demands things no World interface supplies: stable call-site identifiers
and \passthrough{\lstinline!callIndex!} generation, canonical argument
hashing, exposure of replay's next suspension intent, enforcement of the
D1--D3 determinism profile, epoch-chain resolution in place of a single
event store, and new event types with transcript semantics. Those are
\textbf{compiler and runtime changes}, not adapter code: the current
Workflow transform emits step identifiers and suspension hooks, but
nothing like the full intent contract of §7.1. The accurate version of
the headline promise is therefore: \emph{workflow source code unchanged;
the SDK, compiler output, and execution engine beneath it are not.} The
claim stands or falls on an adapter implementing the exact, versioned
contract plus the compiler/runtime extensions above, validated against
the upstream conformance suite -- a named deliverable of §13. What follows
is the shape of the storage-and-queue corner, illustrative only:

\begin{lstlisting}
class P2PWorld implements World {
  private store: Corestore;
  private swarm: Hyperswarm;
  private blobs: TorrentClient;   // BitTorrent v2

  // --- storage: the journal IS a chain of epoch Hypercores (S8.1) ---
  async readJournal(runId: RunId): Promise<AsyncIterable<JournalEntry>> {
    // runId names the genesis epoch; the finalized chain of closure
    // certificates (S6, S8.1) names every epoch after it.
    const chain = await this.resolveFinalizedChain(runId);
    for (const epoch of chain) this.swarm.join(epoch.discoveryKey);
    return decodeCanonicalStream(chain);       // epochs + synthetic epoch.open (A.3)
  }

  async appendJournal(runId: RunId, e: JournalEntry): Promise<void> {
    const head = await this.currentEpochCore(runId);  // finalized chain head only
    if (!head.writable) throw new NotExecutorError(runId);
    await head.append(encode(e));              // signed append; live-replicates
  }

  // --- large payloads: distribution plane ---
  async putPayload(bytes: Uint8Array, enc?: EncryptionPolicy): Promise<PayloadRef> {
    if (bytes.length <= INLINE_MAX) {
      const { keyId, nonce, ct } = await this.keys.envelope(bytes);  // S9.1 - small
      return { kind: "inline", keyId, nonce, ct };                   // buys no exemption
    }
    const { sealed, header } = enc ? seal(bytes, enc) : { sealed: bytes, header: undefined };
    const infohash = await this.blobs.seed(sealed);   // v2 piece tree
    return { kind: "blob", infohash, size: sealed.length, encryption: header };
  }

  // --- queuing: no queue, a swarm of interested wakers (S4.5) ---
  async scheduleWake(runId: RunId, at: HLC): Promise<void> {
    await this.appendJournal(runId, { t: "sleep.begin", seq: next(), wakeAfter: at,
      ...this.rt.currentIntent() });  // (callSite, callIndex, argsHash) - S7.1
    // that's it: the wake obligation is now replicated state,
    // dischargeable by any peer, not a row in one scheduler's table
  }

  // --- authentication: keys and capabilities, not sessions ---
  async assertExecutor(runId: RunId, key: KeyPair): Promise<void> {
    const epoch = await this.finalizedChainHead(runId); // S8.1 - NOT the manifest:
    if (!epoch.executor.equals(key.publicKey))          // the manifest names only the
      throw new NotExecutorError(runId);                // genesis executor; authority
    verifyEpochBinding(epoch, key.publicKey);           // lives in the epoch chain
  }
}
\end{lstlisting}

The essential move is in \passthrough{\lstinline!scheduleWake!}: there
is no scheduler. A sleeping workflow's wake time is journal state,
replicated to every participant. \emph{Anyone} who holds the journal and
cares about the run's progress (the counterparty awaiting the result, a
paid availability peer, the initiator's own infrastructure) can observe
\passthrough{\lstinline!wakeAfter!} pass and initiate resumption. The
centralized queue is replaced by an incentive structure: the parties who
want the workflow to finish are the parties equipped to wake it. §9.2
and §12 treat the incentive question directly.

\hypertarget{the-notarized-resume-flow}{%
\subsection{The notarized resume flow}\label{the-notarized-resume-flow}}

Resumption by a peer other than the original executor is the operation
that makes state ownership genuinely dissolve, so we spell it out.
Suppose executor E crashed mid-run, and participant P (or E's standby,
or a hired availability peer) resumes:

\begin{figure}[H]\centering\includegraphics[width=0.9\linewidth]{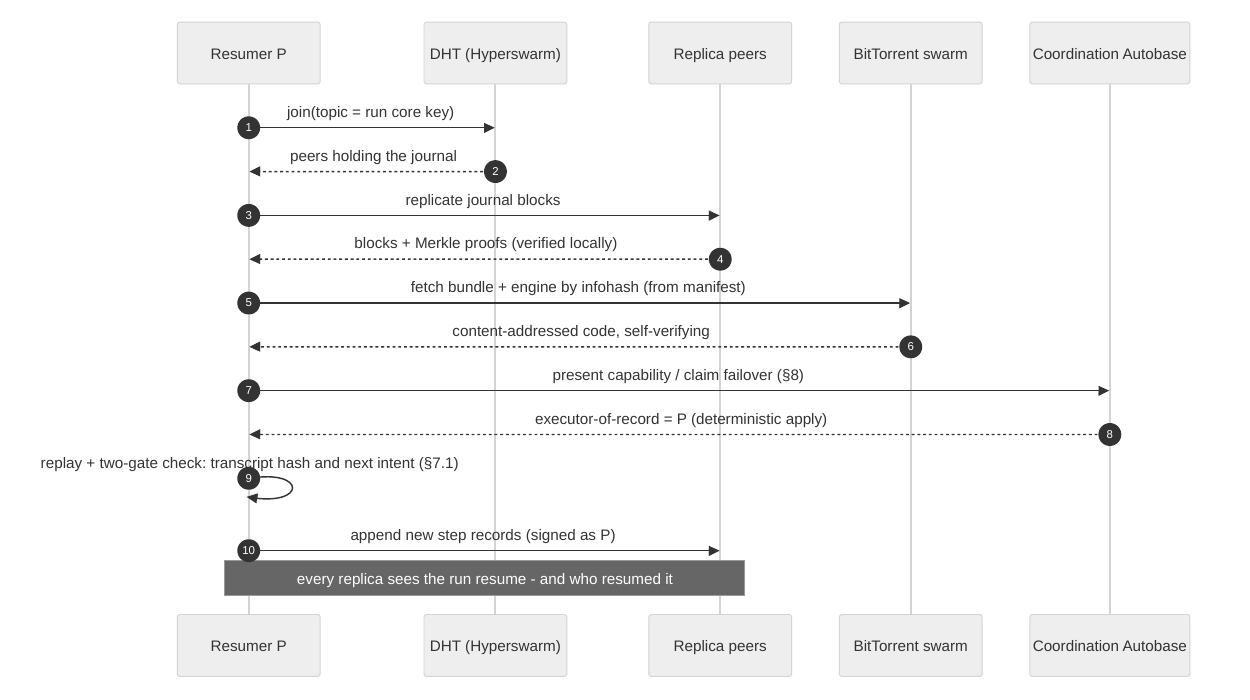}\caption{The notarized resume flow (\S4.5).}\end{figure}

\begin{enumerate}
\def\labelenumi{\arabic{enumi}.}
\tightlist
\item
  \textbf{Fetch.} P joins the DHT topic for the run's core key,
  replicates the journal from any peers holding it, and verifies every
  block against the core's Merkle root. P now holds a history it can
  prove untampered without trusting whoever served it.
\item
  \textbf{Pin code.} P reads
  \passthrough{\lstinline!codeVersion.infohash!} from the manifest,
  fetches the exact workflow bundle from the BitTorrent swarm, and
  verifies it by content address. Replay against the wrong code version
  silently corrupts state (the correlation-ID hazard is well documented
  in self-hosted Workflow SDK deployments); content-addressing the
  bundle in the manifest makes version skew structurally impossible
  rather than operationally avoided.
\item
  \textbf{Authorize.} P's succession takes the only path there is: an
  \passthrough{\lstinline!epoch.open!} in the coordination channel
  (grounded in a voluntary handoff, a fork proof, or a lapsed liveness
  challenge) finalized by the run's indexer majority (§6, §8). P's claim
  carries its readiness declaration (§8.3). The predecessor's key is
  fenced by topology, not by promise: it can no longer extend the run
  (§8.1).
\item
  \textbf{Replay.} P replays the orchestrator deterministically against
  the verified chain, passing both gates of §7.1 (the transcript hash
  and the next-intent comparison) before it may act.
\item
  \textbf{Advance.} P appends new step records to its fresh epoch core.
  Every replica sees the run resume, and sees \emph{who} resumed it, in
  real time.
\end{enumerate}

Five steps, none of which requires an exclusive, liveness-critical
custodian. The run's availability now \emph{tends toward} the investment
of its most interested party (P5's heuristic, with the gaps §9.2 and §12
exist to cover), its integrity rests on cryptography rather than
database ACLs, and its executor succession is an explicit, finalized
chain of closure certificates rather than whoever happens to own the
Redis instance. What these steps do not establish (truth of recorded
outcomes, honesty of any party) is stated in §3 and travels with the run
as explicitly as what they do establish.

\begin{center}\rule{0.5\linewidth}{0.5pt}\end{center}

\hypertarget{design-principles}{%
\section{Design Principles}\label{design-principles}}

\textbf{P1. The journal is the source of truth; the executor is a
cache.} Any design decision that gives the running process authority the
journal lacks is a regression to landlordism.

\textbf{P2. Coordination and distribution are separate planes.} Small
signed facts travel on Hypercores; large immutable bytes travel on
BitTorrent v2. Facts reference bytes by hash; bytes never carry
authority.

\textbf{P3. One run, one writer, always provable.} The single-executor
invariant is enforced by keys, transferred by capabilities, and violated
only with self-incriminating fork proofs.

\textbf{P4. Determinism is the contract, cryptography is the
enforcement.} Deterministic replay makes state a pure function of the
log; signing and Merkle proofs make the log a fact rather than a claim.

\textbf{P5. Availability follows interest, as a heuristic, not a
guarantee.} Every party with a stake in a run replicates it by default;
durability is the sum of stakes, not the SLA of a host. Interest
concentrates \emph{copies}; it does not conjure an online, reachable,
credentialed executor at the moment one is needed, which is why the wake
path (§9.4), failover sets (§8.3), and contracted custody (§12) exist.
Its contrapositive is also policy: a run no one will hold is a run no
one wants finished, and the system lets it die openly rather than
promising universal durability it cannot deliver.

\textbf{P6. Code is content-addressed into the manifest.} A run pins the
exact bundle (and the exact engine) that may replay it. Upgrades are
explicit journal events, not deploy-time accidents.

\textbf{P7. Names and endpoints are advisory; keys are identity.} A run
is its core key. A workflow is its bundle hash. Everything
human-readable resolves through trust policy; binding a friendly name to
a key or a hash is a claim someone signs, never a fact the system
arbitrates.

\textbf{P8. Secrets never enter the replicated plane in clear.} Payloads
are encrypted to participants or to subjects before they touch either
plane (§9.1).

\textbf{P9. Centralized infrastructure may accelerate, never own.} A
vendor can run superpeers, blind replicas, wake services, and capability
gateways -- as competitive services on an open substrate -- exactly as HTTP
seeds accelerate torrents without owning them.

\textbf{P10. Every trust need is met by evidence or by a fungible role.}
Wherever the design seems to need ``someone reliable,'' it must supply
either self-authenticating evidence (fork proofs, transcript
commitments, replication challenges, good behavior made checkable by
anyone) or a role that is stateless and replaceable without moving state
(wake services, blind peers, capability gateways). A component that is
neither is a landlord in embryo and is rejected at design time.

P10 is the organizing principle of the second half of this paper.
Sections 6 and 7 introduce the evidence primitives; §9 shows each hard
problem yielding to evidence, a fungible role, or an explicit admission
of residual risk.

\begin{center}\rule{0.5\linewidth}{0.5pt}\end{center}

\hypertarget{multi-writer-coordination-and-autobase-finality}{%
\section{Multi-Writer Coordination and Autobase
Finality}\label{multi-writer-coordination-and-autobase-finality}}

Single-writer covers the run's spine, but real workflows have inputs
from other parties: approval gates, webhooks, human-in-the-loop hooks,
and (increasingly) other agents. Workflow SDK models these as hooks: the
run suspends awaiting an external payload. In the centralized World, the
hook endpoint is a URL owned by the host; the host's database records
who responded and with what.

In the P2P World, a hook is an Autobase. Each authorized respondent
writes to their \emph{own} Hypercore; Autobase linearizes the
participant cores through its causal DAG and applies them
deterministically. What that machinery provides, and what it does not,
needs saying \textbf{{[}E{]}}: Autobase's documented ordering is
\emph{eventually consistent}: the linearization may reorder as causal
information arrives, and only entries beneath a signed view checkpoint
are stable; checkpoint progress depends on a majority of the designated
indexers continuing to write. Autobase is therefore a
replication-and-linearization mechanism with explicit stabilization
points, not a consensus protocol, and this paper uses it as such.

The design consequences are three rules, applied everywhere the
coordination channel appears \textbf{{[}P{]}}. \emph{Indexer set:} each
run's manifest names its coordination indexers (by default the
participant set, configurably a designated subset) and their majority is
the run's finality quorum. \emph{Finality:} a coordination event is
\textbf{final} when it lies beneath a view checkpoint signed by an
indexer majority; before that it is provisional and may reorder.
\emph{The finality gate:} no irreversible consequence -- an authority
transition (§8), an external effect keyed to an approval, a cancellation
acknowledgment -- may be triggered by a coordination event until that
event is final. Provisional events may drive anticipatory work
(prefetching a bundle, warming a standby) but never authority. A
Byzantine indexer majority can finalize a bad view; that adversary is T3
and out of scope (§3), with BFT indexer sets deferred to §13.

Under those rules, a hook works as follows: the run's
\passthrough{\lstinline!hook.await!} entry names the public keys
entitled to respond and pins the hook Autobase's identity; respondents
append to their own cores; the executor (and every verifier) derives the
hook's outcome by the same deterministic apply, and acts on it at
finality; the responding party's signature travels with their
contribution forever.

The point of the World abstraction is that none of this surfaces in
workflow code. A cross-organizational procurement flow is written just
as it would be against the hosted World; the P2P semantics live entirely
in the hook's grant policy:

\begin{lstlisting}
"use workflow";
export async function procurement(req: PurchaseRequest) {
  const quote = await fetchVendorQuote(req);        // "use step"
  const approval = await hook<Approval>("acme-signoff", {
    grantedTo: [ACME.managerKey, ACME.directorKey], // other org's keys
    quorum: 2,                                      // both must sign
    deadline: "14d",
  }); // suspends; responses arrive on ACME's own cores; the run advances at finality (S6)
  if (!approval.granted) return await notifyRejection(req);  // "use step"
  return await placeOrder(quote, approval.receiptRef);       // "use step"
}
\end{lstlisting}

Under the hosted World, \passthrough{\lstinline!hook(...)!} mints a
callback URL and the host's database records who called it. Under the
P2P World, it appends \passthrough{\lstinline!hook.await!} naming ACME's
keys, and the approval record that unblocks the run is ACME's own signed
appends, jointly held, independently checkable, and never hosted by the
requesting org at all.

One caveat belongs here rather than buried: if ACME's standby sits in
the run's failover set, resuming the run means executing the requesting
organization's step code on ACME's infrastructure; §9.10 names this gap
and the step-ownership discipline that closes it.

The hook model composes into
patterns that are awkward or trust-heavy today.
\textbf{Cross-organizational approval chains:} manager \(\to\) director
\(\to\) VP sign-off where each level is a different org, each approver
appends to their own core under their own key; no org hosts the others'
assent, and the linearized view is a jointly held, independently
checkable approval record by construction. \textbf{Agent-to-agent
contracting:} two agents' negotiation is a pair of single-writer cores
plus an apply function encoding the commitment rules; staged
commitments, escrowed step outputs (encrypted payloads whose keys
release on counterpart signatures), and disputes-with-evidence all fall
out of primitives already described rather than requiring a trusted
intermediary.

Beyond hooks, the run's coordination Autobase serves as its
\textbf{out-of-band channel}, the place where facts about the run that
the executor cannot or will not append are recorded. A silent executor
cannot write its own liveness failure into a single-writer core;
participants record challenges and co-signatures here instead. Sections
7 and 8 lean on this channel for checkpoint co-signing and the eviction
process respectively.

The discipline cost is real and should be stated plainly: Autobase
\passthrough{\lstinline!apply!} functions must be deterministic and must
tolerate reordering before checkpoints. But this is the \emph{same}
discipline durable-workflow authors already accept for orchestrator
code. The P2P World does not import a new mental model; it reuses the
one the SDK already teaches.

\begin{center}\rule{0.5\linewidth}{0.5pt}\end{center}

\hypertarget{transcript-commitments-and-checkpoints}{%
\section{Transcript Commitments and
Checkpoints}\label{transcript-commitments-and-checkpoints}}

A tempting reading of this primitive, that it verifies deterministic
replay by itself, fails for a structural reason: \textbf{a hash of a
fold over the journal is a pure function of the journal.} A resumer that
verifies the journal and recomputes the fold will match the recorded
hash \emph{whatever engine it runs and whatever it is about to do next};
the engine has contributed nothing to the value being checked. The
primitive is therefore named for what it is, a \textbf{transcript
commitment}; what it establishes is enumerated below, and it is extended
with the one component that makes replay itself participate in the
check: the \textbf{next-intent record}.

\hypertarget{the-transcript-commitment-and-the-next-intent-record}{%
\subsection{The transcript commitment and the next-intent
record}\label{the-transcript-commitment-and-the-next-intent-record}}

A \passthrough{\lstinline!transcript.commit!} entry carries two things
\textbf{{[}P{]}}:

\begin{enumerate}
\def\labelenumi{\arabic{enumi}.}
\tightlist
\item
  \textbf{The transcript hash}, the canonical digest of
  \passthrough{\lstinline!TranscriptState!}, the deterministic fold of
  the journal specified in Appendix A. This commits to which steps
  settled with which input/output hashes, what is pending, which hooks
  are open against which Autobase commitments, the manifest in force,
  and the entropy cursor.
\item
  \textbf{The next-intent record}, the executor's declaration, derived
  from its \emph{live continuation} at a suspension boundary, of the
  next command its replay will issue:
  \passthrough{\lstinline!\{ kind: step | sleep | hook | child | end; callSite: bytes; callIndex: number; argsHash: Hash \}!},
  where \passthrough{\lstinline!callIndex!} is the deterministic ordinal
  of this call in the orchestrator's call sequence and
  \passthrough{\lstinline!argsHash!} digests the canonically encoded
  arguments. Unlike the transcript hash, this value is \emph{engine
  output}: it is a function of the bundle, the transcript, \textbf{and
  the engine's actual reconstruction of the continuation.}
\end{enumerate}

A resumer performs both checks before its first append: it recomputes
the fold and compares the transcript hash (catching transcript-decoding
and schema divergence), then replays the orchestrator and compares its
\emph{own} derived next intent against the recorded one (catching
precisely the class the fold alone cannot: an engine that agrees on the
transcript but is about to make a different call). Divergence at either
gate is a \passthrough{\lstinline!TranscriptDivergence!}: fail-detected,
pre-append, with engine pinning (P6) as the recovery path: re-execute
under the engine build named in the last commitment, which reproduces
the recorded intent by construction if the fault was engine skew, and
localizes a genuine nondeterminism bug if not.

\textbf{What this mechanism does and does not establish.} It \emph{does}
establish \textbf{{[}P{]}}: that resumer and committer decode and fold
the same transcript under the same schema; that the resumer's replay
reaches the same next call site with the same arguments before it is
allowed to act; and that checkpoints and receipts commit to a
well-defined state. It does \emph{not} establish: that the two engines'
continuations agree beyond the next call (divergence further along
surfaces only at the \emph{next} boundary check); that recorded step
outputs reflect reality (per §3, log integrity is not truth, and a
malicious executor's false \passthrough{\lstinline!step.end!} passes
every check here); or anything at all about a T2 executor's honesty.
Stronger execution attestation (signed receipts from effect targets,
TEE-attested replay, application-level proofs) is future work (§13). The
name reflects this narrower guarantee.

\begin{lstlisting}
// Two gates against the last commitment, then intent-checked consumption
// of the uncommitted suffix, before the first append. [P]
async function verifiedResume(chain: EpochChain, wc: WorldContract) {
  const entries = await verifyAndDecode(chain);       // Merkle-checked, closure-linked (S8)
  const cm      = lastCommitment(entries);            // commits to prefix [0, cm.seq)
  const prefix  = entries.slice(0, cm.seq);
  if (!equal(transcriptHash(foldTranscript(prefix, wc), wc), cm.stateHash))
    throw new TranscriptDivergence(cm, "transcript"); // gate 1: schema/fold
  const bundle = await fetchByInfohash(manifestOf(entries).workflow.codeVersion.infohash);
  // Replay reads a ReplaySource (A.8), never raw entries: JournalSource here;
  // a pruned failover replica passes CapsuleSource(snapshot, capsule, suffix).
  const replay = engineOfChoice.replay(bundle, new JournalSource(prefix)); // D1-D3
  if (!intentEqual(replay.nextIntent(), cm.nextIntent))
    throw new TranscriptDivergence(cm, "intent");     // gate 2: engine at the boundary
  for (const e of entries.slice(cm.seq)) {            // uncommitted suffix, one op at a time
    if (isIntentBearing(e) && !intentEqual(replay.nextIntent(), intentOf(e)))
      throw new TranscriptDivergence(e, "suffix-intent"); // intentOf total here (S7.1)
    replay.consume(e);                                // advance replay past e
  }
  return replay.resume();                             // first append lands in a fresh epoch
}
\end{lstlisting}

\textbf{The commitment lifecycle, pinned down.} The two-gate check is
well-defined only if the compared commitment matches the replay prefix,
so the protocol pins the correspondence down \textbf{{[}P{]}}: (1)
\emph{Coverage}: a \passthrough{\lstinline!transcript.commit!} at stream
position \emph{p} commits to the fold of positions
\passthrough{\lstinline![0, p)!} and to the next intent of a replay of
that prefix; the entry annotates the suspension (or cadence point)
immediately preceding it and must precede any subsequent operation
entry. (2) \emph{The uncommitted suffix}: entries after the last
commitment (an executor can always crash between a
\passthrough{\lstinline!step.end!} and its next commitment) are consumed
\emph{operation-by-operation under the same intent check}: because every
\emph{intent-bearing} entry carries
\passthrough{\lstinline!(callSite, callIndex, argsHash)!} (the
classification below), every such entry is itself an intent record, and
the resumer verifies that its replay expects exactly that operation
before consuming it; the boundary commitment is the anchor, the suffix
is a checked continuation of it, and a mismatch anywhere is a
\passthrough{\lstinline!TranscriptDivergence!} naming the exact entry.
(3) \emph{Failover from a suffix}: permitted precisely when the suffix
validates; the successor's readiness declaration (§8.3) names the
transcript hash and intent \emph{as of the closure certificate's
\passthrough{\lstinline!branch.treeRoot!}}, uncommitted suffix included.
(4) \emph{Pinned-engine fallback}: a \textbf{recovery attempt}
and not a guarantee. Re-executing under the engine build named in the
last commitment reproduces the recorded intent \emph{if} the divergence
arose from engine skew; if the leak was ambient nondeterminism (a D2
escape), the pinned build can diverge from its own past behavior too, at
which point the run halts fail-detected with a minimizable bug report
against the determinism profile -- the honest floor, and the reason the
profile's conformance corpus (§13.7) matters.

\textbf{Which entries are intents.} For
\passthrough{\lstinline!intentOf(e)!} to be well-defined, the schema
must say which entries carry the tuple, and it does \textbf{{[}P{]}}:
the \textbf{intent-bearing} entries
(\passthrough{\lstinline!step.begin!},
\passthrough{\lstinline!sleep.begin!},
\passthrough{\lstinline!hook.await!},
\passthrough{\lstinline!child.spawn!}, and
\passthrough{\lstinline!run.end!}) each carry
\passthrough{\lstinline!(callSite, callIndex, argsHash)!} (for
\passthrough{\lstinline!run.end!}: kind \passthrough{\lstinline!end!},
\passthrough{\lstinline!callSite!} the completion point's bundler id,
\passthrough{\lstinline!argsHash!} the hash of the canonical result
commitment), and suffix consumption performs the intent comparison on
these and only these. The \textbf{result-bearing} entries
(\passthrough{\lstinline!step.end!},
\passthrough{\lstinline!step.error!},
\passthrough{\lstinline!hook.receive!},
\passthrough{\lstinline!child.join!}) and the \textbf{meta} entries
(\passthrough{\lstinline!transcript.commit!},
\passthrough{\lstinline!checkpoint!},
\passthrough{\lstinline!manifest.ack!}, the synthetic
\passthrough{\lstinline!epoch.open!}) carry no intent and trigger no
comparison; the resumer consumes them as data (cache values, absorbed
errors, view pins, transitions). \passthrough{\lstinline!intentOf(e)!}
is total over intent-bearing entries and undefined by construction
elsewhere, and the suffix-consumption loop in
\passthrough{\lstinline!verifiedResume!} guards on that classification
and nothing else.

Commitments also serve as evidence \emph{against} the executor under T2:
an executor that commits a transcript and next intent and later acts
inconsistently with them has signed material for the dispute, and
combined with fork proofs the honest-executor assumption is replaced by
an audit trail at every suspension point -- an audit trail, not a proof of
honesty.

\hypertarget{checkpoints-and-pruning-verification-is-not-resumption}{%
\subsection{Checkpoints and pruning: verification is not
resumption}\label{checkpoints-and-pruning-verification-is-not-resumption}}

One distinction carries this subsection: \textbf{a commitment can prove
history; it cannot supply history.}
\passthrough{\lstinline!TranscriptState!} collapses settled outcomes
into \passthrough{\lstinline!stepChain!}, a hash accumulator, which is
what Appendix A.1 says it is (``a commitment, not a snapshot'') and why
a peer holding only \passthrough{\lstinline!TranscriptState!} plus the
journal suffix can \emph{verify} the run but cannot \emph{replay} it:
deterministic replay consumes the historical step outputs, hook
resolutions, and tape values themselves, and a hash of them returns
nothing. Checkpointing therefore produces \textbf{two distinct objects
{[}P{]}}, and replica roles are defined by which they retain:

\begin{itemize}
\tightlist
\item
  The \textbf{verification snapshot}: the canonically encoded
  \passthrough{\lstinline!TranscriptState!} (small, O(active work)).
  Sufficient to verify the suffix, serve proofs, and audit; insufficient
  to resume.
\item
  The \textbf{replay capsule}: the compacted, replay-relevant projection
  of the settled prefix (Appendix A.8): every settled operation's
  \passthrough{\lstinline!(seq, callSite, callIndex, argsHash, outcome)!}
  in canonical order (with small outcomes inline and large ones as
  \passthrough{\lstinline!PayloadRef!}s into the distribution plane)
  plus resolved hook outcomes, the consumed entropy/time tape segment,
  and the manifest amendment chain. This is what replay actually reads.
  Its size is O(settled operations) with a small constant (values and
  refs, stripped of signatures, Merkle overhead, and wire framing).
  Pruning therefore bounds \emph{verification} cost at O(active work),
  while \emph{failover} cost remains proportional to the capsule.
\end{itemize}

What a checkpoint co-signer signs, and what it must do before signing,
is a trust boundary worth stating in full, because after pruning the
run's history is exactly as trustworthy as this step. Signing two hashes
endorses two byte strings; it does not establish that both derive from
the same authenticated prefix, that capsule outcomes agree with
\passthrough{\lstinline!stepChain!}, that the snapshot's active state
matches the position, or that capsule-based replay reaches the committed
intent. And ``archival replay will catch a lying quorum later'' cannot
be the primary rule, both because blocks may be cleared after finality
and because an archival replica is not guaranteed to exist. The
obligation is therefore normative \textbf{{[}P{]}}: each honest
cosigner, \emph{before} signing, (1) verifies the journal prefix against
its authenticated root; (2) independently folds the
\passthrough{\lstinline!VerificationSnapshot!}; (3) independently
\emph{derives} the canonical \passthrough{\lstinline!ReplayCapsule!}
bytes via \passthrough{\lstinline!buildCapsule!} (A.8) and compares
hashes (merely byte-verifying executor-supplied bytes is not derivation;
a cosigner that cannot rerun the constructor must verify a proof
equivalent to that derivation or must not sign); (4) replays through
\passthrough{\lstinline!CapsuleSource(snapshot, capsule, emptySuffix)!};
(5) verifies the committed next intent; and (6) signs a typed
certificate binding all of it:

\begin{lstlisting}
type CheckpointCertificate = {
  runId: RunId; epoch: EpochId; branch: BranchId; length: number;  // the prefix
  worldContract: Hash;             // fold + encoding version (Appendix A)
  snapshot: Infohash; capsule: Infohash;
  stateHash: Hash; nextIntent: NextIntent;   // what replay must reach
  constructorVersion: number;      // buildCapsule version (A.8)
  cosigners: PublicKey[]; signatures: Signature[];
};
\end{lstlisting}

One rule shared by all four certificate types (closure, timeout,
acknowledgment, checkpoint): every required signature is over the same
canonical encoding (A.5) of the certificate body \emph{excluding the
signature arrays themselves}, so cosigners sign identical bytes and
signature aggregation cannot change what was attested.

Pruning is licensed only by a finalized
\passthrough{\lstinline!CheckpointCertificate!}, and resume-time trust
then has its plain formulation: \emph{cross-consistency was verified by
the checkpoint quorum before anything was pruned}; a Byzantine
checkpoint quorum is the stated T3 failure (§3), with archival replay
demoted to defense-in-depth where archival replicas happen to exist.
Without this obligation the design would drift, without announcement,
from independently verifiable history to quorum-trusted checkpoint
material at every pruning event; with it, that change is a declared,
certificate-shaped trust boundary. Three replica classes follow
\textbf{{[}P{]}}: \textbf{verification replicas} hold snapshot + suffix
(auditors, dashboards, most participants); they verify and serve proofs
but cannot resume. \textbf{failover-capable replicas} hold
\textbf{snapshot + capsule + suffix}: the capsule supplies settled
replay values, but the \emph{snapshot} is what carries the active state
that is in neither capsule nor suffix (an operation opened before the
checkpoint and still pending at it: an outstanding step, an open hook, a
sleep obligation, a live child, plus the
\passthrough{\lstinline!stepChain!} accumulator, epoch position, and
entropy cursor), and its O(active work) size makes retaining it nearly
free. \textbf{Archival replicas} hold the authenticated genesis history
and remain the only class that can re-derive everything from first
principles.

The Hypercore mechanics themselves are simple \textbf{{[}E{]}}:
\passthrough{\lstinline!truncate!} rewrites (suffix removal, fork bump),
\passthrough{\lstinline!clear!} reclaims local storage only, and there
is no native jointly-authorized prefix discard. A finalized checkpoint
licenses \passthrough{\lstinline!clear()!} below the certificate's
\passthrough{\lstinline!length!} (one term, one meaning:
\passthrough{\lstinline!length!} is the position in the canonical run
stream (§8.1), global across epochs, never epoch-local), per the peer's
chosen class. Fresh verification peers bootstrap from snapshot + suffix,
fresh failover peers from snapshot + capsule + suffix (§7.2's
composition: the snapshot is not optional for resumption), each object
verified against the finalized
\passthrough{\lstinline!CheckpointCertificate!}. Certificate
Transparency's checkpoint trust model is the inspiration here; its log
structure is not imported.

\begin{figure}[H]\centering\includegraphics[width=0.98\linewidth]{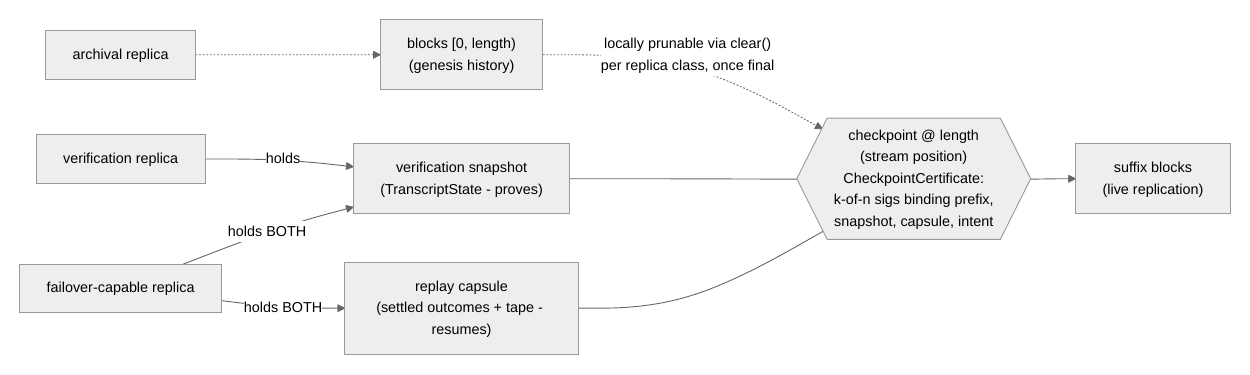}\caption{Checkpoints: verification snapshots, replay capsules, and per-class pruning (\S7.2).}\end{figure}

\hypertarget{checkpoints-as-receipts-for-the-transcript-not-the-world}{%
\subsection{Checkpoints as receipts for the transcript, not the
world}\label{checkpoints-as-receipts-for-the-transcript-not-the-world}}

A finalized checkpoint, carried into a shared attestation log (the
outlook of §11), is a portable receipt with a precise meaning:
\emph{this indexer quorum verified that workflow bundle B, folded over
history H, reaches state S with next intent I.} That is a receipt for a
\textbf{transcript}, disputable and compact, with its verifiability
stated precisely: the object hashes, quorum signatures, and chain
linkage are independently verifiable by anyone; the snapshot--capsule
\emph{semantic} consistency behind them is quorum-attested (§7.2),
independently re-derivable only while un-pruned history exists. It is
not a receipt for the \emph{world}: whether H's recorded effects
occurred is the very gap §3 names, and a receipt consumer who needs
effect-truth needs effect-target attestation (§13). Within that stated
scope, the receipt remains what agent-to-agent coordination lacks today:
proof of process, jointly held.

\begin{center}\rule{0.5\linewidth}{0.5pt}\end{center}

\hypertarget{executor-lifecycle-epochs-fencing-eviction}{%
\section{Executor Lifecycle: Epochs, Fencing,
Eviction}\label{executor-lifecycle-epochs-fencing-eviction}}

The executor role is the one concentration of authority the architecture
retains, and the gap at its center deserves a direct statement: no
capability record in another log can stop a key from signing, so
``transferring append authority'' by fiat is not a mechanism
\textbf{{[}E{]}}. If handoff means giving the successor the run core's
secret key, the predecessor retains it forever; if it means anything
else, run identity and log continuity must be respecified. Fork
detection is post-facto evidence, not revocation. This section therefore
rebuilds the lifecycle around a structure that makes fencing real, and
states plainly which of its guarantees hold under which assumptions of
§3.

\hypertarget{epoch-cores-fencing-by-construction}{%
\subsection{Epoch cores: fencing by
construction}\label{epoch-cores-fencing-by-construction}}

A run's journal is not one Hypercore but a \textbf{chain of epoch cores}
\textbf{{[}P{]}}. Epoch \emph{e} is a single-writer core whose signing
key belongs to that epoch's executor alone; the run is the concatenation
of finalized epochs. An epoch closes and its successor opens through
exactly one mechanism: an \passthrough{\lstinline!epoch.open!} record in
the coordination channel, taking effect \textbf{at coordination
finality} (§6), whose payload is a \textbf{closure certificate}, because
\emph{a length does not identify a history}: an equivocating core has
two different signed trees at the same length, so closing ``epoch 4 at
length 812'' leaves the canonical branch undefined at the exact moment
it matters most. The certificate commits to the authenticated root
\textbf{{[}P{]}}:

\begin{lstlisting}
// Stable identity vs. branch identity - keying safety
// decisions on a (coreKey, forkId) pair conflates them: an EPOCH is stable
// across every branch its key ever signed; a BRANCH is one fork lineage.
type EpochId  = { runId: RunId; epochNumber: number; coreKey: CoreKey };
type BranchId = { forkId: number; treeRoot: Hash };

type EpochClosure = {
  closes: EpochId;                  // the stable epoch - NOT a branch
  branch: BranchId;                 // the branch this closure selects
  finalLength: number;
  lengthProof: Proof;               // signed-length proof for (finalLength, branch.treeRoot)
  successor: CoreKey;               // fresh epoch-core key
  executor: PublicKey;              // successor's principal identity (S4.3)
  basis: HandoffBasis | ForkBasis | TimeoutBasis;   // S8.2, S8.3
  justification: AckCertificate | null; // must equal the highest finalized
                                        // certificate for `closes` whenever one
                                        // exists, on ANY branch, for ANY basis
                                        // (S8.4; reducer below)
  extensionProof: MerkleConsistencyProof | null; // REQUIRED when closing beyond
                                        // a certified root: authenticates that
                                        // branch.treeRoot at finalLength extends
                                        // the certified (treeRoot, length) -
                                        // carried in the closure, because the
                                        // reducer is a deterministic fold and
                                        // must never fetch journal blocks
  readiness: ReadinessDeclaration;  // S8.3
};
\end{lstlisting}

For a cooperative close, \passthrough{\lstinline!branch.treeRoot!} is
the head the predecessor signs off on (its
\passthrough{\lstinline!forkId!} the core's current fork identifier).
For a detected fork, two principles compete. Closing at the \textbf{last
common authenticated ancestor} blesses neither equivocated branch, but
the moment a failover-survivable acknowledgment (possibly
effect-bearing) exists beyond the divergence, ancestor-closure orphans
acknowledged state, and preserving that state blesses a branch. Both
cannot hold, and the design says which yields: \textbf{durability
outranks neutrality {[}P{]}}. The normative fork-closure rule:

\begin{enumerate}
\def\labelenumi{\arabic{enumi}.}
\tightlist
\item
  If no valid \passthrough{\lstinline!AckCertificate!} (§8.4) covers any
  entry beyond the last common authenticated ancestor (the greatest
  length at which both signed trees prove the same tree hash) close at
  the ancestor. Neutrality is free when nothing acknowledged is at
  stake.
\item
  If valid certificates exist beyond the ancestor, §8.4's signer rules
  make them possible on at most one branch. Close \textbf{on that
  branch}, at no less than the highest certified
  \passthrough{\lstinline!(length, treeRoot)!}; the closure carries that
  certificate as its \passthrough{\lstinline!justification!}, and
  closure validation requires the closure's
  \passthrough{\lstinline!branch.treeRoot!} to extend the certified
  root.
\item
  Entries beyond the closure on either branch remain evidence, and any
  effect recorded in them is a possibly-fired orphan under §8.6's
  reconciliation discipline. The equivocator is evicted by the same fork
  proof either way; branch selection changes only one thing:
  acknowledged state is never sacrificed to punish them.
\end{enumerate}

Why this does not open an ``equivocate, then steer'' attack: the
certificate is not the executor's to mint. It exists only if the indexer
ack quorum granted it; honest signers acknowledge monotonically along
one chain (§8.4), so by the time an executor equivocates, whichever
branch holds certificates is already fixed by what was durably
acknowledged. The equivocator chooses where to fork; the protocol (via
durability that predates the fork's detection) chooses what survives.
Readers, verifiers, and the resume flow (§4.5) follow the finalized
chain of closure certificates and nothing else.

Two bookkeeping consequences make the fold well-defined. First,
\passthrough{\lstinline!epoch.open!} lives in the coordination Autobase,
not the journal, so the transcript is redefined as a pure fold over the
\textbf{canonical run stream}: the concatenation of finalized epochs'
journal entries, with each finalized
\passthrough{\lstinline!epoch.open!} inserted as a \emph{synthetic
authenticated entry} at a formally specified position (after the first
\passthrough{\lstinline!finalLength!} entries of the epoch it closes,
i.e., following the prefix \passthrough{\lstinline![0, finalLength)!},
and before the successor's first entry; Appendix A.3). The stream is
deterministic given the finalized coordination view, which is what the
fold is actually a function of. Second, the manifest must bootstrap
coordination unambiguously:
\passthrough{\lstinline!RunManifest.coordination!} carries the Autobase
bootstrap key, the indexer set, and the quorum size, without which
``finalized'' would dangle.

Fencing is then a property of the data structure rather than a promise
about keys: the predecessor's key remains capable of signing appends
\emph{to its own closed epoch core}, but those appends lie beyond the
finalized closing length; every reader following the chain sees them as
what they are -- post-close writes by a fenced writer, evidence rather
than state. Nothing the old key signs can extend the run. This is the
fencing-token pattern of classical distributed systems, materialized as
log topology, and Hypercore's own multi-writer signer manifests are a
candidate implementation substrate for the same effect within one core;
the epoch-chain formulation is chosen here because its safety argument
is inspectable without appeal to library internals, and validating
either construction is a named prototype task (§13). Two consequences
are inherited by everything above: \passthrough{\lstinline!runId!} is
the \emph{genesis} epoch core's key (the chain's root, not the current
writer; §4.3's identity model), and transcript folding (Appendix A)
runs over the finalized concatenation, with each
\passthrough{\lstinline!epoch.open!} entering the fold as a settled
record.

Fencing the journal does not by itself fence the \emph{world}: during a
partition, a stale executor may still fire external effects it believes
it is authorized to make. For capability-native targets, capabilities
are epoch-scoped and expire with the epoch. For legacy targets fronted
by a capability gateway (§9.1), the gateway checks the caller's epoch
against the finalized chain head before exercising the credential, and
this must be named for what it is \textbf{{[}P{]}}: the gateway becomes
a \emph{fencing authority} and a liveness-critical dependency for the
credentials it fronts -- a deliberate, disclosed re-centralization at the
legacy boundary, kept fungible (P10) but not kept trustless. Where
neither epoch-scoped capabilities nor a gateway apply, §8.6's
idempotency-and-reconciler discipline is the remaining bound on
stale-executor damage.

\begin{figure}[H]\centering\includegraphics[width=0.85\linewidth]{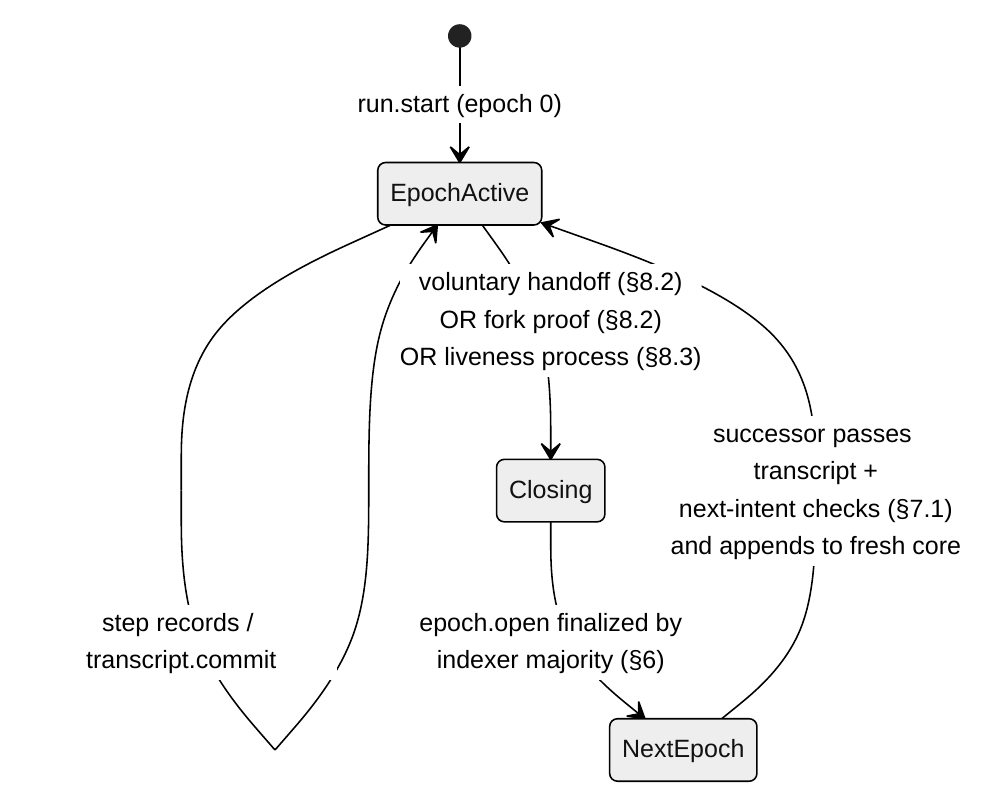}\caption{The executor lifecycle as an epoch chain (\S8). The predecessor key can still sign after the close, but only past its finalized length: fenced by topology, not by promise.}\end{figure}

\hypertarget{closing-an-epoch-voluntary-handoff-and-safety-eviction}{%
\subsection{Closing an epoch: voluntary handoff and safety
eviction}\label{closing-an-epoch-voluntary-handoff-and-safety-eviction}}

\textbf{Voluntary handoff} is the cooperative close: the executor
appends a final \passthrough{\lstinline!transcript.commit!},
co-initiates \passthrough{\lstinline!epoch.open!} naming its successor,
the KEK rotates to the successor's key set (§9.1), and the successor's
first act is the two-gate check of §7.1 across the seam. Planned
rotations (deploys, maintenance, jurisdictional moves) reduce to this.

\textbf{Safety eviction} responds to equivocation: two signed blocks at
one sequence number within an epoch. The fork proof is
self-authenticating \textbf{{[}E{]}} (it cannot be forged against a key
that never double-signed) so any participant presenting one to the
coordination channel initiates an epoch close with no challenge period,
subject only to finality of the resulting
\passthrough{\lstinline!epoch.open!}. It is tempting to call this path's
attack surface empty; the accurate statement is narrower
\textbf{{[}P{]}}: the \emph{trigger} cannot be falsely invoked, but the
surrounding surface remains: a compromised executor key equivocates as
the executor and is evicted as the executor (correctly, but the honest
operator is evicted too); eviction closes an epoch and never adjudicates
which recorded outcomes were true (§3); and a Byzantine indexer majority
finalizing a malicious \passthrough{\lstinline!epoch.open!} is the T3
case the design does not defend.

\hypertarget{liveness-eviction-under-partial-synchrony}{%
\subsection{Liveness eviction under partial
synchrony}\label{liveness-eviction-under-partial-synchrony}}

Silence is not evidence of failure. In an asynchronous network, an
unresponsive executor may be dead, partitioned, censored, or slow, and
no witness set can distinguish these \textbf{{[}E{]}}. The design
therefore does not claim to \emph{prove} liveness failure; it defines a
\textbf{policy decision under the partial-synchrony assumptions of §3},
engineered so that a wrong decision is recoverable rather than
corrupting.

The process: a wake horizon passes unserviced; a challenger records a
liveness challenge in the coordination channel; the executor moots it
with a signed pong or fresh \passthrough{\lstinline!transcript.commit!}.
If the window lapses, the lapse must be \emph{evidenced}, not merely
asserted, and Autobase finality cannot carry that weight alone:
stabilizing an \passthrough{\lstinline!epoch.open!} proves the indexers
ordered it, not that a majority independently watched the clock run out,
and neither the deterministic reducer (which must not read local clocks)
nor an author-controlled HLC (worthless under T2) can substitute. The
liveness basis is therefore a \textbf{timeout certificate {[}P{]}}: a
quorum of independent, signed clock observations:

\begin{lstlisting}
type TimeoutAttestation = {
  challengeHash: Hash;         // binds to one specific challenge
  indexer: PublicKey;          // must be in manifest.coordination.indexers
  firstObserved: HLC;          // when THIS indexer first saw the challenge
  measuredElapsed: number;     // >= window, by this indexer's own monotonic
                               // timer since firstObserved, during which it
                               // observed no valid response
  signature: Signature;
};
type TimeoutBasis = { challenge: Hash; attestations: TimeoutAttestation[] };
\end{lstlisting}

Each attestation asserts two facts from the indexer's own frame, and
both are needed: \emph{when it first observed this challenge}, and
\emph{that its own monotonic timer subsequently measured at least the
window with no valid response observed}. The challenge author's
timestamp plays no role in validity -- a backdated opening buys the
challenger nothing -- because every indexer's window starts when
\emph{that indexer} first saw the challenge. The reducer validates the
certificate structurally: at least \passthrough{\lstinline!quorum!}
attestations from distinct configured indexers, each with
\passthrough{\lstinline!measuredElapsed!} at least the manifest's
window, and mutual \passthrough{\lstinline!firstObserved!} skew within
the manifest's declared bound as a sanity check; Autobase finality then
\emph{stabilizes} the certificate rather than substituting for it. Named
for what it is \textbf{{[}P{]}}: an \emph{attested timeout under the
indexer trust model}, not a cryptographic proof that physical time
elapsed -- a defensible and sufficient claim for a mechanism whose
output is an authorization basis, not a fault proof.

A successor's claim carries a \textbf{readiness declaration},
deliberately not called a proof, because it proves little
\textbf{{[}P{]}}:

\begin{lstlisting}
type ReadinessDeclaration = {
  stateHash: Hash;          // transcript hash as of the closure's branch.treeRoot
  nextIntent: NextIntent;   // intent at that same prefix, uncommitted suffix included
  successor: CoreKey;       // fresh epoch-core key (NOT a principal key - S4.3)
  signature: Signature;
};
\end{lstlisting}

Both values are visible to any replica of the run, so a dishonest or
nonfunctional successor can copy and sign them without replaying
anything. The declaration therefore establishes \emph{identity and
agreement with public values}: not possession of decrypted history, not
a successful replay, not access to effect credentials, not the ability
to remain online or to seed. Under T1 (crash-only, the design's target)
that is operationally sufficient: a crashed standby signs nothing, and
the declaration filters misconfigured successors that would immediately
fail the two-gate check, which is the failure the mechanism exists to
catch. Under T2 it filters nothing, and a manifest that needs more can
set \passthrough{\lstinline!readiness: "challenge"!} to require a
\textbf{readiness challenge}: a fresh nonce and a randomly selected
range from a challenger-chosen object of the finalized checkpoint bundle
(snapshot or capsule, since failover capability requires both, §7.2),
answered with a keyed MAC,
\passthrough{\lstinline!rk = blake3.deriveKey("p2p-world readiness v0", KEK)!},
response
\passthrough{\lstinline!= blake3.keyed(rk, nonce || objectId || rangeSpec || blake3(plaintextRange))!},
binding the challenge, the object and range identity, and a digest of
the plaintext, under a key only KEK-holders can derive. What it
demonstrates should be said precisely: \emph{possession, at response
time, of the plaintext of the challenged range} (equivalently, of the
ciphertext plus the decryption key), which is a different and weaker
claim than either ``holds all history'' or ``can replay.'' That is an
optional policy dial, not a default; and even it attests possession, not
future liveness. Capability to \emph{stay up} remains unproven, which is
why failover-set membership is a vetting decision (§9.2) and not a
theorem, and why eviction-to-a-dead-successor remains bounded by the
same machinery that bounded the original eviction: the next liveness
challenge.

\begin{lstlisting}
// Coordination reducer over FINALIZED events only (S6). [P]
function apply(finalized: LinearizedNode[], view: CoordView, m: RunManifest) {
  for (const n of finalized) switch (n.value.t) {
    case "fork.proof":
      if (verifyForkProof(n.value)) view.recordFork(n.value);       // S8.2
      break;
    case "liveness.challenge":
      view.openChallenge(n.value); break;
    case "liveness.pong":
    case "transcript.echo":
      view.closeChallenge(n.value); break;
    case "epoch.open": {
      const cl = n.value as EpochClosure;
      // GATE 0 -- canonical epoch binding. cl.closes is proposer-supplied
      // and therefore untrusted: the reducer DERIVES the expected epoch
      // from finalized state and rejects anything else. Without this gate
      // a proposal claiming { epochNumber: 700, coreKey: K } occupies a
      // fresh safety-key namespace and bypasses I1 and I3.
      const eid = view.currentEpoch();     // from the finalized chain: runId
                                           // from the channel itself, number =
                                           // predecessor + 1, coreKey = the
                                           // finalized current epoch core
      if (!equal(cl.closes, eid)) break;   // never the proposal's label
      if (view.hasFinalizedSuccessor(eid)) break;         // I1: per EPOCH -- one
                                                          // successor total, however
                                                          // many branches exist
      if (!verifyLengthProof({                            // I2 -- the proof must
            coreKey: eid.coreKey,                         // bind THIS epoch's key
            forkId: cl.branch.forkId,                     // to THIS exact signed
            length: cl.finalLength,                       // (fork, length, root)
            treeRoot: cl.branch.treeRoot,                 // state
            proof: cl.lengthProof })) break;
      // I3 -- for EVERY basis (timeout closures caused the original loss
      // scenario, not just forks): the closure must honor the highest
      // finalized AckCertificate for this EPOCH -- searched across ALL
      // known branches, so a certificate on fork0 constrains a closure
      // that names fork1.
      const cert = view.highestFinalizedAck(eid, applicableConfigs(m, view));
      if (cert) {
        if (!verifyAckCertificate(cert, m)) break;        // threshold, config, sigs (S8.4)
        if (!equal(cl.justification, cert)) break;        // must carry the selector
        // S8.1 rule 2: the selected branch must PROVABLY extend the
        // certified root. A signed-length proof for B only shows the key
        // signed B; an equivocating key signs non-extending roots too.
        const exact = cl.finalLength === cert.length &&
                      equal(cl.branch.treeRoot, cert.branch.treeRoot);
        if (!exact && !verifyConsistency(eid.coreKey,     // Merkle consistency:
              cert.branch.treeRoot, cert.length,          // certified prefix ...
              cl.branch.treeRoot, cl.finalLength,         // ... is a prefix of
              cl.extensionProof)) break;                  // the selected root
      } else if (isForkBasis(cl.basis)) {
        if (!closesAtAncestor(cl, view.forkEvidence(cl.closes))) break; // S8.1 rule 1
      }
      if (!basisValid(cl.basis, view, m)) break;  // basis incl. S8.3 cert
      if (!view.eligibleSuccessor(cl.successor,
                cl.executor, m.policy.failover)) break;
      if (!readinessValid(cl.readiness, view, m)) break;  // S8.3
      view.appendEpoch(cl);                               // fenced at root
    }
  }
}
\end{lstlisting}

What is claimed for this mechanism, precisely \textbf{{[}P{]}}: a valid
timeout certificate is \emph{quorum-observed non-response} -- the
\emph{authorization basis} for an epoch close, not a proof the executor
failed. Under partition, a healthy executor can still be evicted (its
pongs may not have reached the attesting indexers); the design's safety
property is that false eviction is bounded, not impossible: the closure
certificate pins the exact authenticated prefix, the old key cannot
extend the run, in-flight external effects are bounded by epoch-scoped
capabilities, gateway fencing, and §8.6's discipline, and the wrongly
evicted executor holds its own timely pong as dispute material. Floors
on windows, challenge cooldowns, and the
\passthrough{\lstinline!liveness/standard-0!} defaults carry over
unchanged, as does the asymmetry justifying floors-not-ceilings. The
acceptable false-eviction rate is a per-manifest policy trade against
stuck-run duration, measured rather than assumed in §13.6.

\hypertarget{durability-certificates-and-the-finality-quorum}{%
\subsection{Durability certificates and the finality
quorum}\label{durability-certificates-and-the-finality-quorum}}

The ack dial (§9.6), taken alone, permits a loss mode: an executor
running \passthrough{\lstinline!ack: "local"!} appends an outcome,
partitions, and is falsely evicted; the closure certificate, constructed
in good faith from what the indexers can see, closes the epoch at an
earlier length, and the acknowledged entry is orphaned from the
canonical chain. Cryptographically nothing is corrupted; operationally,
acknowledged workflow state is lost, and if the orphan carried an
external effect, the successor may repeat it. Closing that mode requires
more than a flag in the manifest schema asserting that acknowledgments
``intersect the finality quorum'': it requires a protocol whose
guarantees follow from stated signer rules. This section specifies that
protocol \textbf{{[}P{]}}.

\textbf{The acknowledgment certificate.} Failover-survivable durability
is a signed, verifiable object, not a count:

\begin{lstlisting}
type AckCertificate = {
  epoch: EpochId;                   // the stable epoch (S8.1)
  branch: BranchId;                 // the branch lineage and exact root acknowledged
  length: number;                   // covers the prefix [0, length)
  config: Hash;                     // ConfigDescriptor in force (single or joint, S8.4)
  signers: PublicKey[];             // distinct; drawn from the indexer set
  signatures: Signature[];
};
\end{lstlisting}

\textbf{Signer rules, from which the guarantees follow.} (i) \emph{One
universe:} survivable-ack signers are drawn from the run's coordination
indexer set; replica identities and finality identities are the same
universe, so intersection arithmetic is well-defined rather than
assumed. (ii) \emph{Monotone prefix discipline:} an honest signer signs
at most one \passthrough{\lstinline!treeRoot!} per
\passthrough{\lstinline!(epoch, length)!} and only roots that extend
every root it has previously signed in that epoch; an honest indexer
never acknowledges conflicting histories. (iii) \emph{Thresholds:} with
\passthrough{\lstinline!N = |indexers|!} and \passthrough{\lstinline!q!}
the finality quorum, a valid certificate requires
\passthrough{\lstinline!a!} distinct signers where, normatively,

\begin{lstlisting}
a  =  max( ceil((N+1)/2) ,  N - q + 1 )
\end{lstlisting}

The second term gives \passthrough{\lstinline!a + q > N!}: every
certificate intersects \textbf{every possible} finality quorum, so at
least one signer of any closure knows the certified prefix; the
closure-side check of §8.1 has a concrete witness by construction. The
first term gives \passthrough{\lstinline!2a > N!}: any two certificates
share an honest signer under T1, and by rule (ii) that signer cannot
have signed conflicting roots: \textbf{certificates on two conflicting
branches are impossible}, which is what makes the fork rule of §8.1
well-defined. Under T2 with up to \passthrough{\lstinline!f!}
equivocating signers the impossibility requires
\passthrough{\lstinline!2a - N > f!},
i.e.~\passthrough{\lstinline!a > (N+f)/2!}; a manifest targeting T2
durability sets \passthrough{\lstinline!a!} accordingly and accepts the
availability cost. \emph{Canonical epoch identity everywhere:}
\passthrough{\lstinline!AckCertificate.epoch!} and
\passthrough{\lstinline!CheckpointCertificate.epoch!} are subject to the
same rule as closures: an honest signer signs only the epoch identity it
derives from the finalized chain (number = predecessor + 1, key = the
finalized current core, run = the channel's own), and verifiers reject
any certificate naming an epoch that finalized state does not produce; a
proposer-selected epoch label is never a key into the safety state. (iv)
\emph{Verification:} the reducer checks signer distinctness, membership
in the coordination configuration named by
\passthrough{\lstinline!config!}, threshold, and signatures. (v)
\emph{Ordering -- the part that intersection alone does not give:} an
entry \textbf{becomes} survivable only when its certificate is
\textbf{finalized} in the coordination view.
\passthrough{\lstinline!ackDurable("survivable")!} resolves at that
finality, not at signature collection, so by the time the effect
discipline (§8.6) permits a fire, every subsequently finalized closure
deterministically \emph{sees} the certificate, and the race (signatures
collected, effect fired, closure at an earlier root finalizes first,
certificate arrives too late to save the entry) cannot occur: the
certificate either finalizes first and constrains every later closure,
or the ack never resolved and the effect never fired. The alternative
(signers durably remembering signed roots and refusing to finalize
contradicting closures) was considered and rejected for v0: it
introduces a voting rule into finality through extra-protocol signer
state, where the finality-wait keeps the reducer a pure function of the
finalized view. A more accurate account of the cost:
\passthrough{\lstinline!"survivable"!} is not ``an indexer round-trip''
but a \emph{coordination-finality round-trip} -- signature collection plus
time-to-checkpoint -- the very latency §13.4 measures.

\textbf{Reconfiguration.} Membership change is where quorum-intersection
arguments tend to break, so it is joint by rule. One tempting activation
rule, ``the transition ends when a \emph{new-configuration-alone}
certificate finalizes,'' asks a not-yet-valid configuration to authorize
its own validity; the sequence below avoids that circularity by having
every transition authorized by the configuration that is currently valid
\textbf{{[}P{]}}: (1) the indexer-set amendment is a manifest amendment
(§10.3), and its journaling puts the run in the \textbf{joint state}:
finality requires quorums in the old \emph{and} new sets, and survivable
certificates require thresholds in both; (2) the amendment's own
\passthrough{\lstinline!manifest.ack!} is the \textbf{configuration
barrier}, and it is certified \textbf{jointly}: an
\passthrough{\lstinline!AckCertificate!} whose
\passthrough{\lstinline!config!} names the joint descriptor and whose
signer set meets \emph{both} constituent thresholds, which
simultaneously demonstrates that the new set can independently clear its
own threshold; (3) \emph{finalization of that jointly certified barrier
activates the new configuration} -- the joint authority, which is valid,
authorizes its own retirement; (4) certificates after the barrier use
the new configuration alone. A joint configuration is a first-class
object, not a hash pun: \passthrough{\lstinline!config!} names a
canonical
\passthrough{\lstinline!ConfigDescriptor \{ mode: "single" | "joint"; sets: \{ members: PublicKey[]; finalityQuorum: number; ackThreshold: number \}[] \}!}
(finality and acknowledgment thresholds are different numbers and are
carried separately per constituent set) and certificate verification
under a joint descriptor checks signer distinctness and the
\emph{relevant} threshold (finality for view checkpoints,
\passthrough{\lstinline!ackThreshold!} for acknowledgment certificates)
per constituent set. No new entry type is needed; the amendment
certifies its own arrival, under the authority that exists, not the one
being born.

\textbf{The rule and the dial.} In certificate terms:

\begin{quote}
An entry is \textbf{failover-survivable} only if covered by a valid
\passthrough{\lstinline!AckCertificate!}. Closure certificates must
close at no less than the highest certified prefix, on the certified
branch (§8.1), and closure validation rejects anything less.
\end{quote}

The ack dial becomes
\passthrough{\lstinline!ack: "local" | \{ replicas: number \} | "survivable"!}.
\passthrough{\lstinline!ack: "local"!} buys hosted-World latency and
\emph{means} something stated: unprotected against executor loss and
false eviction. \passthrough{\lstinline!\{ replicas: k \}!} buys crash
durability (k Ring-1 replicas hold the bytes) but no closure protection:
a k-replicated entry can still be orphaned by a legitimate fork closure,
because ordinary replicas are not certificate signers.
\passthrough{\lstinline!"survivable"!} is the only level invariant I3
protects, and it costs a coordination-finality round-trip (§8.4 rule v);
effect-bearing steps default to it (intent durable \emph{before} fire,
§8.6, is only as strong as the durability it waits for). This is
invariant I3.

\hypertarget{the-lifecycle-invariants}{%
\subsection{The lifecycle invariants}\label{the-lifecycle-invariants}}

Four invariants now carry the section's safety argument, each with its
enforcing mechanism, all of them prototype-validation targets (§13.3):

\begin{itemize}
\tightlist
\item
  \textbf{I1. Unique succession.} An epoch has at most one finalized
  successor. \emph{Enforced:} the reducer's first check; finality makes
  the winner unambiguous.
\item
  \textbf{I2. Unique history.} A finalized epoch prefix identifies
  exactly one authenticated history. \emph{Enforced:} closure
  certificates commit to \passthrough{\lstinline!branch.treeRoot!} with
  a signed-length proof; a fork closes at the last common ancestor or on
  the certified branch per §8.1, either way at exactly one root, keyed
  by stable \passthrough{\lstinline!EpochId!}, never by branch.
\item
  \textbf{I3. Durable means survivable.} Every failover-survivable
  acknowledgment survives every permitted epoch transition.
  \emph{Enforced:} certificate thresholds make conflicting-branch
  certificates impossible under the stated signer model (§8.4); the fork
  rule preserves the certified branch (§8.1); closure validation rejects
  any closure below, or off, a finalized certificate -- in the reducer,
  for every closure basis -- with the certificate finalized before the
  entry it covers is treated as durable (§8.4 rule v).
\item
  \textbf{I4. Fence before fire.} No executor initiates an external
  effect until its epoch is the finalized chain head and the effect's
  intent record is durable under the step's declared ack policy.
  \emph{Enforced:} the §8.6 ordering discipline, with intent-trace
  fields on \passthrough{\lstinline!step.begin!} (§7.1) making
  violations evident in the record.
\end{itemize}

\hypertarget{the-half-fired-step-external-effects-at-the-seam}{%
\subsection{The half-fired step: external effects at the
seam}\label{the-half-fired-step-external-effects-at-the-seam}}

One hazard lives right at the epoch seam and deserves its own treatment.
Durable execution gives steps at-least-once semantics: if the executor
crashes \emph{after} an external effect fires but \emph{before}
\passthrough{\lstinline!step.end!} is durably appended, the retry
re-fires the effect. Centralized engines share this hazard, but a single
executor with a colocated store keeps the uncertainty window tiny and
private; cross-domain failover widens it: the successor was not the
process that made the call and has no local residue to consult. The
journal, however, already contains the disambiguating fact:
\passthrough{\lstinline!step.begin!} is the intent record. The rule is a
strict ordering discipline plus deterministic idempotency:

\begin{lstlisting}
// Effect discipline: intent -> durable -> fire (keyed) -> outcome.
const idem = blake3.deriveKey("p2p-world idempotency v0", encode({ runId, seq, attempt }));
await world.appendJournal(runId, { t: "step.begin", seq, step, inputRef, attempt,
  callSite, callIndex, argsHash });                 // intent tuple - S7.1
await world.ackDurable(runId, seq);          // per-manifest ack policy (S9.6)
const out = await callExternal(api, input, { idempotencyKey: hex(idem) });
await world.appendJournal(runId, { t: "step.end", seq, step,
  outputRef: await put(out), attempt });
\end{lstlisting}

A successor that finds \passthrough{\lstinline!step.begin!} without a
matching \passthrough{\lstinline!step.end!} knows precisely one thing
(the effect \emph{may} have fired) and, because the idempotency key is
derived from \passthrough{\lstinline!(runId, seq, attempt)!} rather than
from executor-local state, the successor can safely re-issue the
identical call: an idempotent or deduplicating target returns the
original outcome, and the seam heals \textbf{{[}P{]}}. For targets that
support neither idempotency keys nor safe re-query, the manifest's step
policy must declare a \textbf{reconciler} (a read-side probe or
compensation the successor runs before retrying) or mark the step
\passthrough{\lstinline!at-most-once!}, accepting that a seam crash
surfaces as an explicit \passthrough{\lstinline!step.error!} for the
workflow to handle rather than a silent double-fire. Capability gateways
additionally keep a bounded dedupe window over recent keys, which under
§8.1 doubles as their fencing check. What this section refuses to do is
pretend the seam away: exactly-once external effects do not exist in any
World, centralized or not \textbf{{[}E{]}}, and this architecture's
contribution is making the uncertainty window a replicated, inspectable
fact instead of a private one, under the explicit external-effect
assumptions of §3.

\begin{center}\rule{0.5\linewidth}{0.5pt}\end{center}

\hypertarget{hard-problems-proposed-resolutions-and-their-limits}{%
\section{Hard Problems: Proposed Resolutions and Their
Limits}\label{hard-problems-proposed-resolutions-and-their-limits}}

Each subsection states the proposed resolution first and its limits
last; per §1, these are \textbf{{[}P{]}} designs pending the validation
of §13, not settled results.

\hypertarget{secrets-and-confidential-state}{%
\subsection{Secrets and confidential
state}\label{secrets-and-confidential-state}}

The problem splits into \textbf{data confidentiality} and \textbf{effect
authority}, which are easy to conflate. Data confidentiality is
well-trodden territory, but its sharp edges require explicit treatment:
envelope encryption per payload, with data-encryption keys (DEKs)
wrapped to a run-scoped key-encryption key (KEK) held by the participant
set and rotated at every epoch transition (§8.1). Crucially, DEKs are
scoped \textbf{per subject} (per data-subject or per counterparty) not
per run, which §9.3 will spend. The journal replicates ciphertext and
hashes only; keys travel the delegation chain, never the log.
Replayability is thereby capability-scoped: any peer can verify
integrity; only key-holders can reconstruct plaintext and execute -- a
feature for confidentiality, a sizing constraint for the failover set
(your failover set is your key-holding set, and the manifest now
declares both).

Effect authority (the credentials with which steps act on external
systems) is the genuinely hard half, because centralized Worlds solve it
silently with ambient IAM that does not travel with a journal. Two
mechanisms, by target type. For capability-native services, the manifest
carries UCAN-style attenuated delegation chains: the principal delegates
``call API X, scope Y, until T'' to the executor role, and any
authorized resumer presents the same chain; authority becomes portable
because it was made explicit. For legacy services with static bearer
credentials, we accept a service, but a P10-compliant one: the
\textbf{capability gateway}, a minimal fungible proxy that holds exactly
one raw credential and exercises it for requests signed by the current
holder of the run's executor capability, verifying the delegation chain
per call. The gateway holds a secret but never state; it can be swapped,
self-hosted, or run redundantly without touching the run; it is to
credentials what a blind peer is to bytes, and under §8.1 it is
additionally the fencing authority for the targets it fronts -- a
disclosed liveness-critical role. \emph{Residue:} the gateway is a
confidentiality and availability dependency for the credentials it
fronts, and minimizing the trust in it (threshold-splitting the
credential, TEE attestation of the proxy) is left open as an engineering
track (§13) rather than baked into the core design.

\textbf{The confidentiality perimeter, specified {[}P{]}.} Four points
that are easy to leave implicit, made explicit here. \emph{Inline
payloads:} the \passthrough{\lstinline!inline!} variant of
\passthrough{\lstinline!PayloadRef!} is enveloped the same way as blobs
(\passthrough{\lstinline!\{ kind: "inline", keyId, nonce, ct \}!}); small
size buys no exemption from P8. \emph{The journal itself:} entry
\textbf{structure} (types, step names, call sites, argument hashes,
timing) is sensitive metadata, so journal cores use Hypercore's block
encryption as an explicit ring boundary: Ring 1 admission grants
replication of \emph{encrypted blocks} (lengths, positions, and timing
visible, contents not) while the block key is Ring 2 material held by
keyed participants. Confidentiality of entry structure against a Ring-1
replica is thus a stated property, not an accident; what Ring 1 still
leaks (traffic timing, entry sizes, participation) is the §9.8/§13.7
privacy residue. \emph{Argument hashes:} a raw
\passthrough{\lstinline!argsHash!} over low-entropy arguments invites
dictionary confirmation by anyone who sees it, so the hash is keyed per
run, with the context slot and the key slot used for what they are:
\passthrough{\lstinline!argsKey = blake3.deriveKey("p2p-world args v0", runSecret)!}
derives a per-run subkey from run key material under a fixed public
context, and
\passthrough{\lstinline!argsHash = blake3.keyed(argsKey, canonicalEncode(args))!}
MACs the canonical arguments under it, verifiable by keyed parties,
opaque to everyone else, and no secret ever rides in a context string.
\emph{Rotation and rewrap:} KEKs are per participant-set epoch;
membership change or epoch transition mints a new KEK and appends rewrap
records for live DEKs. Rewrapping is forward-looking only: parties keyed
for past data retain what they could already read (§9.3's frankness
about erasure applies to revocation identically), and the
forward-secrecy limits of membership change are those stated in §9.8.

\hypertarget{availability-tiers-proofs-and-honest-death}{%
\subsection{Availability: tiers, proofs, and honest
death}\label{availability-tiers-proofs-and-honest-death}}

No token economy. Availability is a three-tier structure declared in the
manifest's \passthrough{\lstinline!liveness!} field, plus one
verification primitive. \textbf{Mutual-stake} (the default for
multi-party runs): participants replicate by protocol requirement; the
counterparty awaiting your output is thereby equipped and motivated to
hold and wake the run, and P5 does real work with no further mechanism.
\textbf{Contracted:} a commodity market of blind peers (replicas holding
encrypted cores they cannot read, a pattern the Hypercore ecosystem
already ships) compensated out-of-band, with compensation gated on
\textbf{proof-of-replication challenges}: any participant issues a
random block-range query, and a peer that cannot answer against the
Merkle root within the response window has failed to demonstrate
possession -- evidence of breach under the custody proof's stated timing
assumption (§12.2), not an unconditional proof. Cheap to run, hard to
fake under those assumptions, and requiring no consensus machinery:
evidence again, per P10. \textbf{Best-effort:} the run lives while
anyone cares. And the design's candid position is that this tier's
failure mode is correct behavior: a run that reaches zero replicas is a
run every stakeholder abandoned, and encoding mortality outright beats
simulating a durability landlord. \emph{Residue:} none at the mechanism
level -- the contracted tier's settlement design is §12's custody
market; what stays open there is anti-outsourcing hardening of the
custody proof and venues for price discovery (§13).

\hypertarget{erasure-versus-append-only}{%
\subsection{Erasure versus
append-only}\label{erasure-versus-append-only}}

Crypto-shredding, sharpened by §9.1's key granularity, with the limits
given first, because they are real \textbf{{[}E{]}}: destroying keys
removes \emph{future} access through the replicated planes; it cannot
recall plaintext or keys that authorized participants already copied
outside them. Crypto-shredding is an architecture for not creating new
access plus an auditable record of the erasure contract, not a memory
hole, and no replicated system offers one. Within that scope:
per-subject DEKs mean erasing one person's data destroys \emph{their}
keys only; their payloads become permanently opaque ciphertext hashes
while every other party's data survives. The
\passthrough{\lstinline!redacted!} replay claim also needs its condition
\textbf{{[}P{]}}: a shredded step replays as
\passthrough{\lstinline!redacted!} soundly \emph{only if} the
orchestrator's recorded control flow never observed the plaintext value;
the decision sequence must be reconstructible from journal structure and
surviving outputs alone. Where a shredded value influenced a branch,
redacted replay cannot reproduce the run; the manifest must either scope
such steps out of erasure or accept that erasure ends replayability at
that point, and identifying value-observing control flow at build time
(taint analysis in the bundler) is future work (§13).
\passthrough{\lstinline!RetentionPolicy!} declares the shredding
contract at \passthrough{\lstinline!run.start!} so every peer replicates
knowingly, and the inviolable rule remains: one plaintext payload
gossiped to a swarm is unerasable forever, so P8 is enforced at the
World boundary, not left to application discipline.

\hypertarget{timers-without-a-scheduler}{%
\subsection{Timers without a
scheduler}\label{timers-without-a-scheduler}}

Wake obligations should not require wakers to replicate journals. Each
application publishes a \textbf{horizon core}: a Hyperbee index mapping
\passthrough{\lstinline!(wakeAfter -> runId)!}, appended alongside every
\passthrough{\lstinline!sleep.begin!}. Wake services subscribe to
horizons only (a few bytes per sleeping run) and their job is a
notification, not an execution: on horizon expiry, ping the executor's
key-holders through the coordination channel. If the ping goes
unanswered past grace, it \emph{is} the liveness challenge of §8.3 (an
authorization basis under §3's synchrony assumptions, not a fault proof)
and the epoch machinery takes over; missed wakes therefore delay but
never corrupt, and late arrival finds an unambiguous chain head. Wake
services are the archetypal P10 fungible role (stateless, redundant,
competitive, hireable, fireable) and racing wakers are harmless because
waking confers no authority; only capability-holders append. A vendor's
hosted wake service (P9) is simply the best-in-class competitor in this
market, not its owner.

\hypertarget{determinism-across-heterogeneous-peers}{%
\subsection{Determinism across heterogeneous
peers}\label{determinism-across-heterogeneous-peers}}

Pin \emph{and} verify, with the verification now properly scoped. The
manifest pins the engine build by infohash next to the bundle (P6),
distributable through the same swarm. The transcript commitment's two
gates (§7.1) are the empirical check, and each catches a different
class: the fold comparison catches schema and decoding divergence (and
is engine-independent by design: the fold has nothing engine-specific in
it), while the next-intent comparison is where the \emph{engine's own
replay} is examined, catching an engine about to diverge at the
boundary. What no boundary check catches \textbf{{[}E{]}} is divergence
that first manifests \emph{between} boundaries; the design's containment
for that class is the discipline itself (the D1--D3 profile of A.6) plus
the guarantee that any divergence becomes visible at the next commitment
rather than compounding silently, with engine pinning as a recovery
\emph{attempt} when the profile is caught leaking -- an ambient leak can
defeat the pinned build too (§7.1). Cross-engine agreement is therefore
a \emph{managed property} (profiled, gated at seams, pinned on failure)
not a theorem. \emph{Open:} the conformance corpus and cross-engine
differential harness (§13) are what would turn the profile from design
into evidence.

\hypertarget{latency-and-the-durability-dial}{%
\subsection{Latency and the durability
dial}\label{latency-and-the-durability-dial}}

The structural mitigations stand: journals replicate ahead of need, so
reads are local; appends are local-then-gossiped; the swarm is touched
on first contact with a run, not per step. What the resolution adds is
candor about the write path: a local append on a disk that dies with its
executor was never durable, and centralized Worlds get their durability
ack (database commit) invisibly. The P2P World makes the ack an explicit
per-run dial,
\passthrough{\lstinline!ack: "local" | \{ replicas: number \} | "survivable"!},
whose semantics §8.4 states rather than implies:
\passthrough{\lstinline!ack: "local"!} commits a step on local append
(fastest, trust-your-disk) and is \textbf{not protected against
executor loss or false eviction} -- a locally-acked entry can be
orphaned by a legitimate epoch closure.
\passthrough{\lstinline!\{ replicas: k \}!} commits only when \emph{k}
Ring-1 replicas confirm replication of the block (the real price of
joint custody, paid in one gossip round-trip), buying crash durability
but still no closure protection.
\passthrough{\lstinline!"survivable"!} buys an
\passthrough{\lstinline!AckCertificate!} from the indexer threshold
(§8.4), the only level I3 protects, at the price of a
coordination-finality round-trip (§8.4 rule v); effect-bearing steps
default to it, closing the loop with §8.6, because the ordering
discipline (intent durable \emph{before} fire) is only as strong as the
durability level it waits for. High-stakes steps can vary the dial per
entry. The P2P World still loses microbenchmarks to a colocated
database, and should: it is buying joint custody, verifiability, and
executor portability, which no colocated database sells at any latency;
the ack dial just prices the purchase per run instead of hiding it.

\hypertarget{the-unit-of-swarming}{%
\subsection{The unit of swarming}\label{the-unit-of-swarming}}

Proposed as a default with an escape hatch: runs aggregate into a
\textbf{per-application Corestore swarm} (one discovery topic, coarse
replication, few DHT announcements), and a run gets its \textbf{own
topic} only when its participant set crosses the application's trust
domain -- the cross-org and agent-to-agent cases where a stranger must
discover exactly one run and nothing else. The residual privacy concern
(swarm-topic membership correlates a peer's interests) is handled at two
levels: derived discovery keys already prevent outsiders enumerating
which runs exist, and principals wanting unlinkability front their
participation with blind peers, so the swarm sees the blind peer's
address, not the principal's. Insider correlation (participants
observing each other) is not a defect to engineer away: it is governed
by the participant set, which is the run's declared trust boundary
anyway.

\hypertarget{the-wire-possession-gated-admission-not-acls}{%
\subsection{The wire: possession-gated admission, not
ACLs}\label{the-wire-possession-gated-admission-not-acls}}

Transport security itself is table stakes here: every Hyperswarm
connection is a Noise-encrypted, mutually authenticated secret-stream,
so confidentiality and integrity on the wire come with the substrate.
The design work is above the cipher: \emph{admission} (who may even
connect and replicate) and \emph{metadata} (what the network learns from
watching). The resolution for admission is to have no admission service
at all, but a \textbf{four-ring model in which every ring is gated by
possession of a different secret}, each derivable only from the ring
inside it, so access control is a property of key material rather than
of any server's ACL table:

\begin{figure}[H]\centering\includegraphics[width=0.4\linewidth]{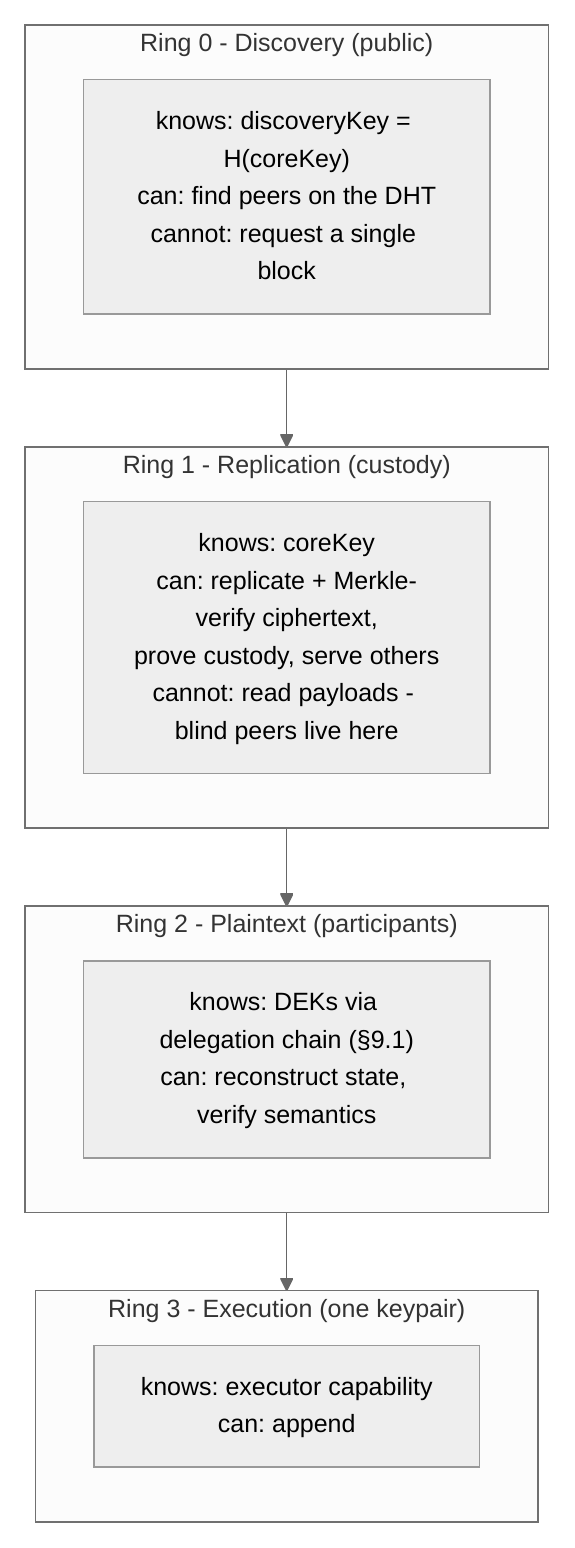}\caption{The four-ring possession-gated admission model (\S9.8).}\end{figure}

The rings are enforced by the primitives themselves. The DHT stores only
the \emph{discovery key} (a one-way hash of the core key) so watching
the DHT reveals that some topic has peers, never which run it is or how
to fetch it. Hypercore's replication handshake requires proof of the
core key itself, so even \emph{ciphertext} is unobtainable by a peer
that merely observed the swarm: replication admission is a capability
check performed by mathematics on every connection, with no revocation
list to serve and no admission server to seize; the landlord-shaped
component that ``private networking'' usually brings with it simply does
not exist. DEKs and the executor capability gate the inner rings as §9.1
and §8 already specify. Membership change composes with §10.3: an
amendment removing a participant rotates the KEK, so removal is
forward-secret: the removed party keeps what it already replicated
(unavoidable in \emph{any} replicated system, and stated rather than
hidden) and can decrypt nothing appended after the rotation. Relayed
connections (when holepunching fails) see ciphertext and endpoint
addresses only, and relays are archetypal P10 fungible roles; principals
wanting endpoint privacy front themselves with blind peers, so the relay
observes blind-peer traffic, not the principal's. \emph{Residue:} a
global passive adversary correlating traffic timing and volume across
swarms defeats all of the above without breaking any of it; cover
traffic and mixnet integration are real countermeasures with real costs,
and this design leaves them as a declared open track (§13) rather than
pretending topic-hashing solves traffic analysis.

\hypertarget{scale-pull-is-backpressure-cores-are-cheap-fan-out-is-fission}{%
\subsection{Scale: pull is backpressure, cores are cheap, fan-out is
fission}\label{scale-pull-is-backpressure-cores-are-cheap-fan-out-is-fission}}

The scale question (millions of runs, thousands of parallel steps)
decomposes into connections, memory, flow control, and hot logs, and
each has a structural answer rather than a tuning answer.

\textbf{Connections do not multiply with runs.} All cores shared between
two peers replicate over a single multiplexed secret-stream (protomux),
and the per-application swarm default (§9.7) means one DHT announcement
and one socket per peer pair carry every run those peers share. A
million runs between two organizations is one connection.

\textbf{Dormant runs cost disk, not RAM.} A cold core is a closed file
until someone requests it; Corestore hydrates lazily. The natural
lifecycle (most workflow runs sleep most of the time) maps onto the
storage tier where sleeping is free, and checkpointed truncation (§7.2)
bounds even the disk.

\textbf{Flow control is the default posture, not a bolt-on.} Hypercore
replication is pull-based: readers request the ranges they want, when
they want them. A slow replica therefore slows only itself; it can never
exert pressure on the executor's append path, which is bounded by local
disk plus whatever \passthrough{\lstinline!ack!} coupling the manifest
\emph{explicitly} purchased (§9.6). Contrast push-based queue
architectures, where backpressure is a failure behavior to engineer;
here it is the protocol's resting state, and the only producer-side
coupling in the entire system is the one the run declared in its
durability policy.

\textbf{Hot runs undergo fission.} The one genuine hotspot is a single
run fanning out to thousands of parallel steps, all contending to append
to one totally ordered log. The resolution is \textbf{child-run cores}:
each parallel branch becomes a child run (its own core, its own manifest
inheriting the parent's policy and participant set) and the parent
journal records only \passthrough{\lstinline!child.spawn(coreKey)!} and
\passthrough{\lstinline!child.join(coreKey, checkpointHash)!}. This is
structured concurrency materialized as log topology: every individual
core stays at a write rate one signer trivially sustains, branch
journals truncate independently once joined, and verification composes:
the parent's transcript commits to its children's checkpoints, so a
verifier descends only into the branches it disputes. The mapping is
also semantic, not just mechanical: durable-execution frameworks already
model this as child workflows, so the fission boundary in the log is the
fission boundary the programming model draws anyway. The same shape
solves horizon load: the per-application horizon core (§9.4) shards into
time-bucketed cores (one per day-bucket), so wake services replicate
only the near future instead of every sleeping run's entire schedule.

\begin{center}\rule{0.5\linewidth}{0.5pt}\end{center}

\hypertarget{foreign-code-on-local-machines-the-isolation-gap}{%
\subsection{Foreign code on local machines: the isolation
gap}\label{foreign-code-on-local-machines-the-isolation-gap}}

One gap in this design deserves its own heading rather than a scattered
treatment, because everything above quietly assumes it away: a peer that
resumes a run fetches the manifest-pinned bundle by infohash and
executes it, and nothing in the architecture isolates that code from the
host. Node.js is not a sandbox and neither is Bare:
\passthrough{\lstinline!vm!} is documented as not a security boundary,
\passthrough{\lstinline!worker\_threads!} share the process, and a
bundle pulled off the swarm runs with the executing peer's filesystem,
network, and environment \textbf{{[}E{]}}. The paper has a determinism
story for code (A.6) and a provenance story for code (§11), but no
isolation story. The vetting argument of §9.2 and the ring model of §9.8
govern \emph{who may execute}, not \emph{what the code can do once it
runs}, and the headline use case is where this bites hardest: in the
cross-org procurement flow of §6, ACME's standby resuming the run means
ACME executes the requesting organization's step code on ACME's
infrastructure. Trusting a counterparty's signature on recorded outcomes
is a very different thing from trusting their code on your machines, and
the two must not be conflated.

The existing mechanisms cover part of the exposure, and it is worth
saying which part. The capability gateway (§9.1) bounds
\emph{credential} exposure: foreign code presents a signed request and
never holds the raw secret. TEE-attested replay (§13.7) is attestation
\emph{to others} that you ran the right code, not protection of the host
\emph{from} the code. Neither touches the host's own filesystem,
network, and keys.

The two halves of the programming model are asymmetric in a way that
points at the resolution \textbf{{[}P{]}}. Orchestrator code under the
D1--D3 profile is pure replay with the ambient surface closed by the
build (A.6): no clock, no randomness, no environment probes, results
delivered in journal order. That is a genuinely sandboxable artifact, in
a V8 isolate or a WASM target, and sandboxing it is an engineering task
rather than a research problem. \passthrough{\lstinline!"use step"!}
functions are the opposite by construction: effectful leaves with full
runtime access, because reaching out to the world is the reason they
exist. They cannot be sandboxed without destroying their purpose, and
OS-level isolation per run (a microVM, gVisor) would protect the host
only by reintroducing the operational surface that the resume-anywhere
argument was built against.

So the problem is precisely the steps, and the proposed resolution is to
give steps an \textbf{owner}: no principal ever executes another
principal's effectful code. The manifest assigns each step an owning
principal; the run's executor executes only the steps it owns; a foreign
step is dispatched exactly like a hook or a child run (§6, §9.9): the
owner executes it on its own infrastructure, appends the signed result
to its own core, and the fold consumes that result at coordination
finality. \emph{Executor-of-the-run} no longer implies
\emph{executor-of-every-step}; failover for a step is scoped to that
step owner's failover set; and the orchestrator remains the only shared
code, which is exactly the sandboxable half. This is more consistent
with P10 than the current design, not less: it replaces a trusted party
who runs everyone's code with evidence, in the form of signed step
results on the owner's own core. The cost is also stated plainly: step
ownership touches the manifest schema (per-step ownership, per-owner
failover sets), changes the meaning of the
\passthrough{\lstinline!failover!} field, and adds a finality round-trip
to every foreign step, and its design and validation are named work in
§13.7.

Until that discipline is in place, the honest floor is this: the
architecture as specified assumes that an executor trusts the code of
every participant whose steps it runs, and prospective failover peers
should read the \passthrough{\lstinline!failover!} field as accepting
exactly that trust.

\hypertarget{what-the-landlord-also-sold-streams-observability-amendments-migration}{%
\section{What the Landlord Also Sold: Streams, Observability,
Amendments,
Migration}\label{what-the-landlord-also-sold-streams-observability-amendments-migration}}

Section 9 proposed resolutions for the hard problems; this section
addresses the conveniences a hosted World bundles alongside custody. A
hosted World bundles four conveniences beyond the journal itself
(durable streaming, observability, run administration, and the comfort
of not migrating) and the P2P World must account for each, because
``decentralized but featureless'' loses to ``centralized but complete''
every time.

\hypertarget{durable-streams-are-native-not-a-feature}{%
\subsection{Durable streams are native, not a
feature}\label{durable-streams-are-native-not-a-feature}}

Hosted Worlds engineer durable streaming (token-by-token AI output
surviving disconnects and deploys) on Redis-backed streams with
reconnection bookkeeping. In the P2P World the entire feature dissolves
into the substrate, because a durable, resumable, verifiable stream is
\emph{what a Hypercore is}. A streaming step appends chunks to a
dedicated stream core and records its key in the journal:

\begin{lstlisting}
const stream = store.namespace(runId).get({ name: `stream-${seq}` });
for await (const token of llm.generate(prompt)) await stream.append(token);
await world.appendJournal(runId, { t: "step.end", seq, step, attempt,
  outputRef: { kind: "stream", core: stream.key, length: stream.length } });
\end{lstlisting}

A client is just a sparse replica: it follows appends live, and
reconnection is not a protocol feature but the absence of one:
replication resumes at whatever block it last held, from \emph{any} peer
carrying the core, with every chunk Merkle-verified. The
reconnectable-client machinery that hosted Worlds advertise is here
indistinguishable from ordinary replication.

\hypertarget{observability-without-privileges}{%
\subsection{Observability without
privileges}\label{observability-without-privileges}}

Hosted observability (run inspectors, traces, time-travel debugging) is
a privileged view over the vendor's database. The P2P World inverts the
access model: an inspector is a \textbf{read-only peer}. The local
\passthrough{\lstinline!workflow inspect!} experience is a replica of
one run's core rendering the same entries the executor wrote; an
org-wide dashboard is an \textbf{indexer peer} that replicates the
application swarm and folds journals into Hyperbee indexes (by workflow,
status, wake horizon, error class) -- a fungible P10 role anyone can run,
race, or replace, since its output is recomputable from the logs. Two
properties fall out that centralized observability cannot offer.
Auditability without access grants: a regulator or counterparty inspects
with full fidelity given only replication and the relevant capability
keys (§9.1): no vendor account, no API quota, no trust in the
dashboard's honesty, since every rendered fact carries its inclusion
proof. And time-travel as a first-class right: any peer can replay to
any committed transcript because that is simply what the data \emph{is}.

\hypertarget{cancellation-and-manifest-amendments}{%
\subsection{Cancellation and manifest
amendments}\label{cancellation-and-manifest-amendments}}

Runs are administered, not just executed: cancelled, re-scoped, their
participant sets changed. Cancellation is a signed coordination event
from an authorized principal, acted on at finality (§6); the executor
acknowledges by appending \passthrough{\lstinline!run.end(cancelled)!},
and an executor that ignores a valid cancel is thereby \emph{silent with
respect to a replicated obligation}, which is the condition §8.3 already
handles, so cancellation needs no new enforcement machinery -- only the
declaration that unacknowledged cancels are challengeable. Amendments
generalize this: participant changes, policy retuning, or failover-set
rotation are co-signed coordination events forming a \textbf{manifest
chain}: each amendment references its predecessor's hash, and verifiers
evaluate every journal entry against the manifest version in force at
its HLC. The run's constitution is thus itself an append-only, jointly
held document, amended by the quorum rules the original manifest
declared.

\hypertarget{migration-the-mirror-world}{%
\subsection{Migration: the Mirror
World}\label{migration-the-mirror-world}}

Nobody adopts a state layer by flag day, and the journal's portability
is precisely what makes gradualism cheap. The adoption path is three
phases of shifting custody, not three rewrites:

\begin{figure}[H]\centering\includegraphics[width=0.95\linewidth]{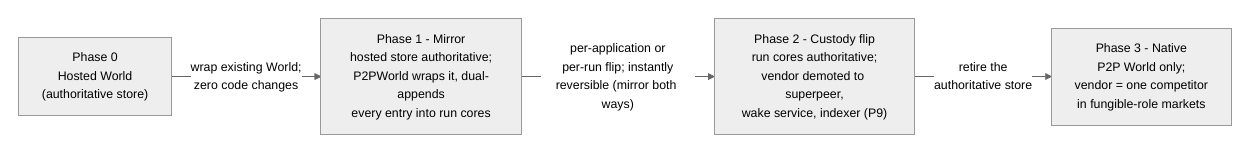}\caption{The Mirror World migration path (\S10.4).}\end{figure}

Phase 1 is implementable today as a \passthrough{\lstinline!World!}
decorator: every \passthrough{\lstinline!appendJournal!} writes through
to the hosted store \emph{and} to a run core, so participants gain
verifiable replicas, receipts, and disaster-grade portability while the
hosted store remains authoritative. The mirror is insurance, and
insurance has premiums, which should be listed rather than waved off:
dual-write latency on the append path, egress and storage cost for the
mirrored cores, key-distribution work before the first encrypted byte
moves, the possibility of dual-write divergence (which the transcript
commitment turns into a detected fact rather than a silent one, but
someone still reconciles it), and the operational surface of running
swarm connectivity at all. The claim is that these premiums are modest
and the coverage (an exit path that does not require the vendor's
cooperation) is structurally unavailable at any price today; the claim
is not that the mirror is cheap to try. Phase 2 flips authority per run
or per application, and is reversible by mirroring in the opposite
direction, because ``export'' and ``import'' have collapsed into the
same operation: replication. That collapse is the strategic point. The
exit cost that makes hosted state a landlordship is an artifact of
proprietary journal custody; once the journal is a core, changing Worlds
is changing who else holds what everyone already holds.

Calling this ``pure upside insurance'' would be too kind: insurance has
premiums, and Phase 1's are concrete \textbf{{[}E{]}}. \emph{Dual-write
divergence:} the hosted store and the run core can disagree (partial
failures, ordering differences), so the decorator needs a declared
authority rule (Phase 1: hosted wins) and a reconciliation job that
detects and reports divergence rather than letting two ``authoritative''
histories drift. \emph{Latency:} every append pays the slower of two
writes, or gives up read-your-writes on one side. \emph{Egress and
storage:} the mirror doubles write bandwidth and adds replica storage,
with cloud egress priced accordingly. \emph{Key distribution:} encrypted
mirroring requires standing up the KEK/DEK machinery of §9.1 before any
custody benefit accrues -- the hardest part of the target architecture,
paid during the phase with the least payoff. Mirror World is
\emph{worthwhile} insurance (the exit-cost asymmetry it buys is real)
but it is bought, not free.

\begin{center}\rule{0.5\linewidth}{0.5pt}\end{center}

\hypertarget{outlook-one-trust-fabric-from-code-to-execution}{%
\section{Outlook: One Trust Fabric from Code to
Execution}\label{outlook-one-trust-fabric-from-code-to-execution}}

Two artifacts in this design want to outlive the participant set that
made them, and both point at the same missing piece. A checkpointed
transcript (§7.3) is a receipt any counterparty can verify, but only if
they can \emph{find} it and trust its context. And a run's
\passthrough{\lstinline!codeVersion.infohash!} (P6) pins exactly which
bytes may replay the journal, but resolves to bytes, not to provenance:
nothing yet says who published that bundle, under what policy, or
whether it was later yanked. The natural completion is an
\textbf{attestation log}: a shared, append-only, Merkle-structured
transparency log (Certificate Transparency's trust model, generalized
from certificates to claims) into which execution receipts, custody
records (§12), and code-publication claims are notarized. This paper
stops short of specifying that log; it specifies the shape of its
citizens. A publication claim would be a manifest of the same species as
the \passthrough{\lstinline!run.manifest!} (identity, content root,
lineage, policy) binding provenance to the infohash a run already pins,
so that every run inherits supply-chain provenance by reference; a
notarized receipt would close the loop from ``this code was published by
this identity'' to ``and it executed this history to this state.'' The
thesis extends by induction: distribute bytes through swarms, record
authority in signed append-only structures no single party owns, and the
same fabric that removed the state landlord removes the registry
landlord above it. That system is future work; nothing in this paper
depends on it, and everything in this paper composes with it.

\begin{center}\rule{0.5\linewidth}{0.5pt}\end{center}

\hypertarget{future-work-an-evidence-native-custody-market}{%
\section{Future Work: An Evidence-Native Custody
Market}\label{future-work-an-evidence-native-custody-market}}

This section is future work on purpose: the market below depends on
mechanisms whose validation is still ahead (coordination finality for
breach co-witnessing, §6; custody timing assumptions, §12.2), and
nothing in the core architecture requires it; the mutual-stake and
best-effort tiers of §9.2 stand alone. It is retained in design form
because the contracted availability tier instantiates the oldest
unsolved question in peer-to-peer systems (\emph{why should a stranger
hold your bytes?}) and because the shape of an answer constrains the
core design (custody contracts consume checkpoints, snapshots, and the
coordination channel as already specified). Three commitments keep the
proposal engineering rather than tokenomics.

\textbf{Commitment 1: money moves only against evidence.} The protocol's
job is to define the evidence objects (offer, contract, challenge,
response, receipt, breach) as signed, replicated facts. Payment is a
function of evidence, never of promise or reputation-in-advance.

\textbf{Commitment 2: settlement is pluggable.} The evidence layer is
rail-agnostic: an epoch receipt can settle by invoice and bank transfer,
by a stablecoin payment channel, by Lightning, or by an escrow contract.
No protocol token exists, because the moment custody requires a bespoke
currency, the currency's operators are the new landlord.

\textbf{Commitment 3: one market for all custody.} A peer holding an
encrypted run journal, a workflow bundle, a model checkpoint, or an
archived history is performing the same service: provable possession of
content-addressed bytes. Custody contracts are therefore written against
\passthrough{\lstinline!(contentId, size, term)!} regardless of what the
bytes mean, so every custody demand this design generates (journals,
payloads, archives, code) draws on a single supply pool, which is what
makes the supply side liquid enough to be cheap.

\hypertarget{the-contract-and-the-loop}{%
\subsection{The contract and the loop}\label{the-contract-and-the-loop}}

\begin{lstlisting}
type CustodyContract = {
  contentId: CoreKey | Infohash;      // journal core OR distribution-plane object
  custodian: PublicKey;
  principal: PublicKey;
  terms: {
    pricePerEpoch: RailAmount;        // rail-typed, not protocol-typed
    epochLength: Duration;
    challengeRate: number;            // expected challenges per epoch
    responseSLO: Duration;            // anti-outsourcing bound, see S12.2
    settlement: RailDescriptor;       // invoice | channel | escrow | ...
    term: { start: HLC; end: HLC };
  };
  signatures: [Signature, Signature]; // both parties; contract is itself evidence
};

// One challenge round. Both messages are posted to the contract's
// coordination channel, so presence and absence are equally public.
type CustodyChallenge = { contractId: Hash; nonce: Nonce;
                          range: BlockRange; issuedAt: HLC };
type CustodyResponse  = { contractId: Hash; nonce: Nonce;
                          proof: MerkleProof;               // blocks in known root
                          binding: Hash;                    // H(nonce || blockBytes)
                          answeredAt: HLC };
\end{lstlisting}

\begin{figure}[H]\centering\includegraphics[width=0.98\linewidth]{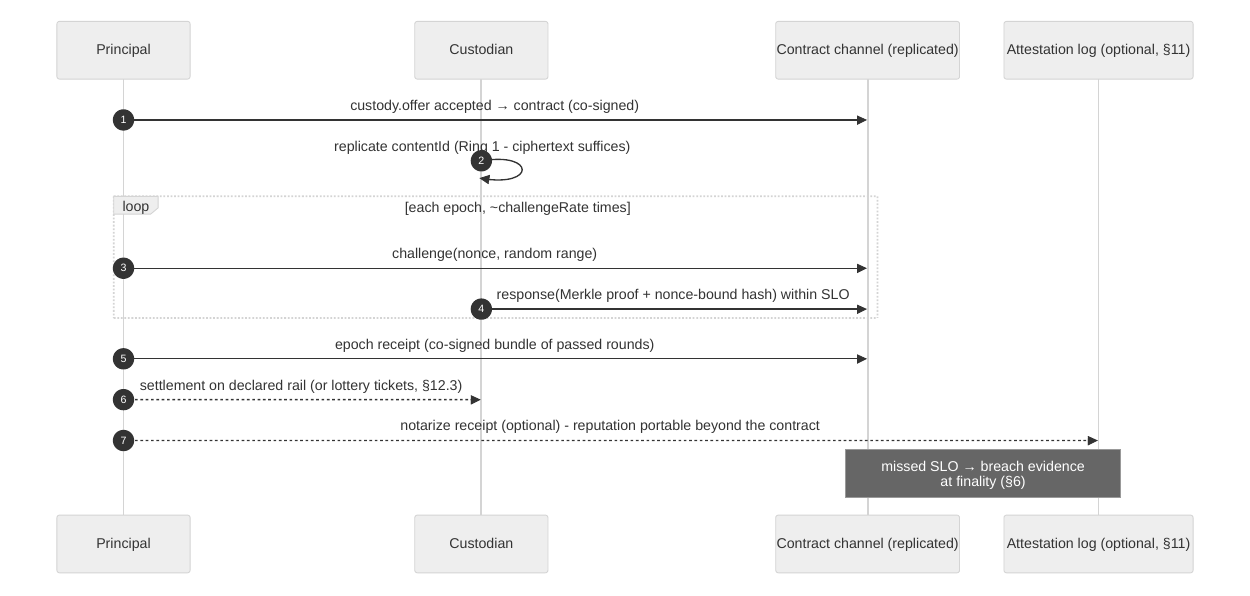}\caption{The proposed custody-market loop (\S12.1, future work). A missed response SLO becomes a timeout claim; quorum-observed non-response at coordination finality (\S6) constitutes breach evidence, leading to non-payment and an optional notarized breach record.}\end{figure}

The loop reuses machinery this paper already built rather than inventing
enforcement. Challenges and responses live in a replicated coordination
channel, so a breach claim rests on \emph{quorum-observed non-response
at coordination finality} (§6) rather than on accusation -- the same
authorization-by-observed-non-response construction as §8.3. The
custodian needs only Ring 1 access (§9.8): it proves possession of
ciphertext against the Merkle root, so the market clears without any
custodian ever being trusted with plaintext.

\hypertarget{the-custody-proof-and-its-stated-assumption}{%
\subsection{The custody proof and its stated
assumption}\label{the-custody-proof-and-its-stated-assumption}}

A Merkle proof alone shows the custodian \emph{can obtain} the blocks,
not that it \emph{holds} them: a fake custodian could fetch from another
peer on demand. Two measures narrow the gap without closing it. The
\textbf{nonce binding}
(\passthrough{\lstinline!H(nonce || blockBytes)!}) forces the responder
to possess the raw bytes at challenge time, not merely a cached proof.
The \textbf{response SLO} makes outsourcing a race: the SLO is set below
the plausible fetch-then-respond latency for the content size, so a
custodian that doesn't hold the data must beat the network to keep its
receipts. This is a timing assumption, and the paper states it as one:
against a custodian colocated with another replica, timing bounds
weaken, and the residual hardening -- challenges over \emph{encodings
unique per contract} (e.g., custodian-sealed replicas, as in
proof-of-replication literature) -- is flagged in §13 as the known upgrade
path rather than baked in, because it buys resistance to a marginal
adversary at real encoding cost, and the base market must stay cheap
enough to use.

\hypertarget{settlement-mechanics-and-reputation}{%
\subsection{Settlement mechanics and
reputation}\label{settlement-mechanics-and-reputation}}

Per-epoch invoicing works today with zero exotic infrastructure (the
receipt is a signed artifact any accounting system can pay against) and
is the default. Where per-challenge granularity matters (short
contracts, adversarial counterparties), \textbf{probabilistic
micropayments} drop in without changing the evidence layer: each
challenge response doubles as a lottery ticket: the response binding,
hashed with a jointly derived beacon, wins
\passthrough{\lstinline!pricePerEpoch!} with probability
\passthrough{\lstinline!1/challengeRate!}, so expected payment equals
the contract price while settlement events per epoch drop to roughly
one, amortizing any rail's fixed costs. Non-payment against a valid
receipt is, symmetrically, principal-side breach: the receipt plus an
unsettled rail state is itself evidence, so reputational consequences
cut both ways.

Reputation is where the market meets the outlook of §11: epoch receipts
and breach records are natural citizens of a shared attestation log,
making a custodian's history (volume held, challenge pass rate,
breaches, counterparty diversity) a queryable, portable,
cryptographically grounded record rather than platform stars. Price
discovery stays, by design, \emph{off-protocol}: offers are gossiped or listed
anywhere, because a mandated venue would be a landlord for the market
itself. The protocol's promise is narrower and stronger: wherever you
found each other, and whatever rail you settle on, the evidence that
determines who owes whom is jointly held and independently checkable,
all the way down.

\begin{center}\rule{0.5\linewidth}{0.5pt}\end{center}

\hypertarget{research-agenda-and-evaluation-plan}{%
\section{Research Agenda and Evaluation
Plan}\label{research-agenda-and-evaluation-plan}}

A proposal should name what would falsify it. The agenda below is
ordered by how much each item bears on the architecture's central
claims; the first three are the ones on which the design stands or
falls.

\textbf{13.1 Minimal prototype.} Build the smallest system that
exercises the core loop, leaving out everything §12 defers: one run as
an epoch chain; a fixed executor and one replicated cold standby; code
and engine bundles pinned by infohash; deterministic replay under the
D1--D3 profile; transcript commitments with next-intent records; and
exactly one carefully specified handoff path (voluntary first, then
lapsed-challenge failover). Executor and standby are the same principal
in this scope, so §9.10's isolation gap does not bite here; it becomes
load-bearing at the first cross-principal failover set. Success
criterion: a standby on a different machine, and then a different engine
build, resumes a suspended run through the two-gate check with zero
manual intervention, and a deliberately skewed engine is \emph{stopped}
by the intent gate.

\textbf{13.2 Prototype requirements for checkpoint composition.} Two
checkpoint-composition details are requirements on the prototype, not
open questions. \textbf{(R1) Dual position coordinates.}
\passthrough{\lstinline!CheckpointCertificate.length!} is the global
canonical-stream position, but a Hypercore tree root is authenticated at
an \emph{epoch-local core length}, and \passthrough{\lstinline!clear()!}
operates on physical per-core coordinates. The prototype must specify
the deterministic mapping, either a
\passthrough{\lstinline!CheckpointPosition \{ streamPosition, epochLength \}!}
pair in the certificate or a normative
\passthrough{\lstinline!projectStreamPositionToEpoch(length, finalizedEpochChain)!},
and the certificate must bind, or allow derivation and verification of,
the current epoch's signed-length proof at that
\passthrough{\lstinline!epochLength!}. Absent this, implementations will
disagree about which blocks a checkpoint covers and which physical
ranges may be cleared. \textbf{(R2) Checkpoint configuration
discipline.} Checkpoints license irreversible local pruning and must
follow the same configuration discipline as acknowledgments:
\passthrough{\lstinline!CheckpointCertificate!} gains
\passthrough{\lstinline!config: Hash!};
\passthrough{\lstinline!ConfigDescriptor.sets[]!} gains a per-set
\passthrough{\lstinline!checkpointThreshold!} with the mechanical bound
\passthrough{\lstinline!1 <= checkpointThreshold <= |members|!} (making
§7.2's \passthrough{\lstinline!checkpointQuorum > f!} rule verifiable
rather than prose on a free scalar); during the joint window a
checkpoint certificate must satisfy the checkpoint threshold in
\emph{both} constituent sets, and only the new set applies after the
jointly authorized barrier. Old-only signers must not be able to license
pruning on the eve of their retirement.

\textbf{13.3 Fencing validation.} Beyond the base fencing cases, the
harness must cover: racing an epoch closure against certificate
finalization (§8.4's ordering rule); closures naming a different branch
than a finalized certificate (§8.1's cross-branch lookup); and
capsule-bootstrapped resume with a pre-checkpoint pending operation.
Implement the epoch chain against real Hypercore semantics and attack
it: a fenced predecessor attempting post-close appends, racing
successors, \passthrough{\lstinline!epoch.open!} under partitioned
indexers, and the alternative construction via Hypercore signer
manifests. The deliverable is a documented safety argument grounded in
observed behavior, or a documented failure that revises §8.1.

\textbf{13.4 Coordination finality measurement.} Characterize Autobase
empirically for this use: time-to-checkpoint versus indexer count and
connectivity; reordering frequency before stabilization; behavior when
indexer majorities pause; and the end-to-end latency the finality gate
(§6) adds to approvals and epoch transitions. The design assumed these
costs are acceptable; the measurements decide.

\textbf{13.5 World conformance.} Implement the exact, versioned
\passthrough{\lstinline!@workflow/world!} contract (queue, stream, and
storage surfaces, hook deduplication, atomicity requirements and all)
and run the upstream conformance suite. Publish the results either way;
``code unchanged, only a new World'' graduates from hypothesis only
here.

\textbf{13.6 Systems measurements.} Local versus k-replica
durability-ack latency (§9.6); resume time as a function of journal
length, with and without capsule bootstrap, and capsule size versus
journal size across workloads (§7.2); split-brain behavior under induced
partitions, including realized false-eviction rates versus policy dials
(§8.3); Corestore overhead at large dormant-run counts (§9.9);
NAT/holepunch success rates in realistic topologies; and the
payload-size crossover between Hypercore-native blobs and BitTorrent v2
distribution.

\textbf{13.7 Open problems} (design unsettled, not merely unvalidated):
\emph{step-ownership execution} (per-owner step dispatch with signed
results consumed at finality, per-owner failover sets, and orchestrator
sandboxing in an isolate or WASM target under D1--D3, closing §9.10's
isolation gap); \emph{effect-truth attestation} (signed receipts from
effect targets, TEE-attested replay, or application-level proofs,
closing the log-integrity-is-not-truth gap of §3);
\emph{Byzantine-tolerant coordination} (BFT indexer sets or external
anchoring for T3 regimes); \emph{custody-proof hardening and market
design} (§12); \emph{erasure-aware control flow} (bundler taint analysis
for §9.3's redaction condition); \emph{the conformance corpus} (test
vectors and a cross-engine differential harness for Appendix A,
including the canonical BitTorrent v2 payload wrapping of §4.3); and
\emph{network-observer privacy} (swarm-metadata analysis per observer
class), the one attacker the four-ring model explicitly does not answer.

\begin{center}\rule{0.5\linewidth}{0.5pt}\end{center}

\hypertarget{conclusion}{%
\section{Conclusion}\label{conclusion}}

Two communities built the same object without noticing. Durable
execution frameworks represent a long-running computation as a
deterministic function of an append-only log; the peer-to-peer community
made append-only logs signed, verified, sparsely replicable, and jointly
held. The defensible thesis of this paper is a narrow one:
\textbf{durable workflow journals are promising objects for
authenticated, multi-party replication, and a P2P-backed World may
provide state portability and independently verifiable custody without
changing workflow source code.}

Around that thesis, this paper has proposed (and marked as proposed) the
machinery a serious attempt needs: a threat model that separates log
integrity from truth; transcript commitments whose next-intent records
make replay itself participate in verification; epoch cores that make
fencing a property of topology rather than a promise about keys; a
coordination layer used strictly within Autobase's documented
guarantees, with explicit finality gates; pruning designed against
Hypercore's real semantics; and an effect discipline that surfaces
at-least-once uncertainty instead of hiding it.

What has the argument actually shown? Something bounded, and worth
stating with care. Exclusive storage custody is not a necessary feature
of durable execution: the journal can be jointly held, independently
verified, and portably resumed, at least in design. Scheduling, fencing,
credentials, availability, and finality all still carry the load. What
this paper does is relocate them -- out of a single operator's private
machinery and into named, inspectable, mostly fungible mechanisms -- and
its obligation, discharged in §13, is to say which experiments would
show those mechanisms carrying the load. The landlord's \emph{storage}
was never load-bearing. Whether everything else the landlord did can be
held up by evidence and fungible roles is the question this architecture
exists to test.

\begin{center}\rule{0.5\linewidth}{0.5pt}\end{center}

\hypertarget{acknowledgments}{%
\section*{Acknowledgments}\label{acknowledgments}}

The authors thank the anonymous technical reviewers of earlier drafts;
their critiques materially improved this design.

\appendix

\hypertarget{the-transcript-encoding-worldcontract-specification-draft-v0}{%
\section{\texorpdfstring{The Transcript Encoding: \texttt{worldContract}
Specification (Draft
v0)}{The Transcript Encoding: worldContract Specification (Draft v0)}}\label{the-transcript-encoding-worldcontract-specification-draft-v0}}

This appendix specifies what a
\passthrough{\lstinline!transcript.commit!} commits to, precisely enough
that two independently implemented engines either produce identical
hashes or fail closed. It is the specification that §7 (attestations,
checkpoints, receipts), §8 (eviction defenses), and §12 (custody
receipts over checkpointed content) stand on.

\hypertarget{goal-and-non-goals}{%
\subsection{Goal and non-goals}\label{goal-and-non-goals}}

\textbf{Goal:} the transcript hash is a \emph{pure function of the
journal},
\passthrough{\lstinline!transcriptHash = H(encode(fold(entries)))!},
computable by any conformant implementation from the verified log alone,
with no reference to any engine's memory layout, scheduler, or suspended
continuations.

\textbf{Non-goals:} the transcript is a \textbf{commitment}, not a
\textbf{snapshot}. It does not aim to contain enough information to
resume execution (the journal itself is the resumption medium, §4.5),
and it excludes, on principle, everything reconstructible from the
journal, because anything included redundantly is a fresh opportunity
for two engines to disagree about representation. The design rule
throughout: \emph{when in doubt, leave it out and let the journal carry
it.} What resumption reads when the journal prefix is pruned is a
different object with the opposite rule -- the replay capsule (A.8) --
and the two must not be conflated: a commitment that tried to double as
a resumption medium would inherit every representational disagreement
the transcript exists to exclude.

\hypertarget{the-transcript-state}{%
\subsection{The transcript state}\label{the-transcript-state}}

\begin{lstlisting}
type TranscriptState = {
  wc: string;               // worldContract version, e.g. "p2p-world/0"
  manifestHash: Hash;       // head of the manifest amendment chain in force (A.4)
  seq: number;              // highest non-meta journal seq folded
  stepChain: Hash;          // running commitment to ALL settled step outcomes (A.2.1)
  pending: PendingStep[];   // open steps only - the active transcript, sorted (A.5)
  sleeps: OpenSleep[];      // unexpired sleep obligations, sorted
  hooks: OpenHook[];        // unresolved hooks with their Autobase commitments (A.4)
  children: ChildRef[];     // spawned/joined child runs (S9.9), sorted by coreKey
  epoch: number;            // finalized epochs folded so far (S8.1)
  entropyCursor: number;    // values consumed from the entropy/time tape (A.9)
  result: Hash | null;      // set only by run.end; terminal
};

type PendingStep = { seq: number; stepId: bytes; inputHash: Hash; attempt: number };
type OpenSleep   = { seq: number; wakeAfter: HLC };
type OpenHook    = { seq: number; hookId: bytes; grantSetHash: Hash;
                     base: { key: CoreKey; length: number; viewRoot: Hash } };
type ChildRef    = { coreKey: CoreKey; state: 0 | 1;  // 0 = spawned, 1 = joined
                     checkpointHash: Hash | null };
\end{lstlisting}

\hypertarget{the-step-chain-history-by-accumulator-not-by-list}{%
\subsubsection{The step chain: history by accumulator, not by
list}\label{the-step-chain-history-by-accumulator-not-by-list}}

Settled outcomes (completed steps, absorbed errors, resolved hooks,
acknowledged handoffs) do not accumulate as a list: a long run would
make the transcript unboundedly large and its encoding a scaling hazard.
They fold into a single running hash:

\begin{align*}
\mathit{stepChain}_0 &= \mathrm{blake3.deriveKey}(\text{"p2p-world stepchain v0 "} + wc,\ \mathit{runId})\\
\mathit{stepChain}_i &= \mathrm{blake3}(\mathit{stepChain}_{i-1} \,\Vert\, \mathrm{encode}(\mathit{settledRecord}_i))
\end{align*}

where $\mathit{settledRecord}_i$ is the canonical encoding
(A.5) of the settling entry's commitment tuple -- for a completed step,
\passthrough{\lstinline!(seq, stepId, inputHash, outputHash, attempt)!}.
The transcript thus stays \textbf{O(active work)}, never O(history): its
variable-size fields are only the pending set, open sleeps, open hooks,
and live children, all of which the programming model already bounds
(they are the run's current concurrency, not its past). Checkpoint
truncation (§7.2) composes trivially: a truncated peer resumes folding
from the checkpointed \passthrough{\lstinline!TranscriptState!}, whose
\passthrough{\lstinline!stepChain!} already commits to everything below
the cut.

\hypertarget{the-fold-a-total-fail-closed-transition-table}{%
\subsection{The fold: a total, fail-closed transition
table}\label{the-fold-a-total-fail-closed-transition-table}}

\passthrough{\lstinline!fold!} is defined by exactly one transition per
journal entry type. Two global rules precede the table.
\textbf{Meta-exclusion:} \passthrough{\lstinline!transcript.commit!} and
\passthrough{\lstinline!checkpoint!} entries are skipped entirely (they
neither mutate state nor advance \passthrough{\lstinline!seq!}) because
an attestation that hashed prior attestations would be self-referential,
and two engines could then disagree about agreement itself.
\textbf{Fail-closed totality:} an entry whose type (or field shape) the
implementation does not recognize is not skipped, not
best-effort-parsed, but a \textbf{fold error} that halts resumption.
Extension therefore \emph{requires} a \passthrough{\lstinline!wc!}
version bump (A.10); there is no such thing as an ``ignorable'' journal
entry, which is precisely how silent cross-engine divergence is made
impossible rather than unlikely.

\begin{longtable}[]{@{}
  >{\raggedright\arraybackslash}p{(\columnwidth - 2\tabcolsep) * \real{0.5000}}
  >{\raggedright\arraybackslash}p{(\columnwidth - 2\tabcolsep) * \real{0.5000}}@{}}
\caption{The fold: a total, fail-closed transition table
(A.3).}\tabularnewline
\toprule\noalign{}
\begin{minipage}[b]{\linewidth}\raggedright
Entry
\end{minipage} & \begin{minipage}[b]{\linewidth}\raggedright
Transition
\end{minipage} \\
\midrule\noalign{}
\endfirsthead
\toprule\noalign{}
\begin{minipage}[b]{\linewidth}\raggedright
Entry
\end{minipage} & \begin{minipage}[b]{\linewidth}\raggedright
Transition
\end{minipage} \\
\midrule\noalign{}
\endhead
\bottomrule\noalign{}
\endlastfoot
\passthrough{\lstinline!run.start!} & initialize state;
\passthrough{\lstinline!manifestHash <- H(manifest)!};
\passthrough{\lstinline!entropyCursor <- 0!} \\
\passthrough{\lstinline!step.begin!} & append to
\passthrough{\lstinline!pending!} (keyed by
\passthrough{\lstinline!seq!}; duplicate key = fold error) \\
\passthrough{\lstinline!step.end!} & remove
\passthrough{\lstinline!pending[seq]!} (absent = fold error); absorb
\passthrough{\lstinline!(seq, stepId, inputHash, outputHash, attempt)!}
into \passthrough{\lstinline!stepChain!} \\
\passthrough{\lstinline!step.error!},
\passthrough{\lstinline!willRetry!} & remove
\passthrough{\lstinline!pending[seq]!}; absorb error tuple into
\passthrough{\lstinline!stepChain!} (the retry arrives as a fresh
\passthrough{\lstinline!step.begin!} with
\passthrough{\lstinline!attempt+1!}) \\
\passthrough{\lstinline!step.error!}, terminal & as above; additionally
absorb terminal marker \\
\passthrough{\lstinline!sleep.begin!} & append to
\passthrough{\lstinline!sleeps!} \\
any entry with \passthrough{\lstinline!hlc > s.wakeAfter!} & first, drop
every such expired \passthrough{\lstinline!s!} from
\passthrough{\lstinline!sleeps!} (expiry is judged by \emph{journal}
HLCs (recorded facts) never wall clocks; A.7) \\
\passthrough{\lstinline!hook.await!} & append to
\passthrough{\lstinline!hooks!} with
\passthrough{\lstinline!grantSetHash!} and the hook Autobase's
\passthrough{\lstinline!(key, length, viewRoot)!} at grant time \\
\passthrough{\lstinline!hook.receive!} & remove
\passthrough{\lstinline!hooks[hookId]!}; absorb
\passthrough{\lstinline!(seq, hookId, payloadHash, from)!} into
\passthrough{\lstinline!stepChain!} \\
\passthrough{\lstinline!epoch.open!} \emph{(synthetic, §8.1)} & inserted
into the canonical stream at its specified position from the finalized
closure certificate; absorb
\passthrough{\lstinline!(closes, branch, finalLength, successor, executor)!}
into \passthrough{\lstinline!stepChain!};
\passthrough{\lstinline!epoch += 1!} \\
\passthrough{\lstinline!manifest.ack!} \emph{(schema addendum)} &
\passthrough{\lstinline!manifestHash <- chainHead!}, coordination-plane
amendments (§10.3) enter the fold only through the executor's journaled
acknowledgment, keeping the transcript a function of one log \\
\passthrough{\lstinline!child.spawn!} /
\passthrough{\lstinline!child.join!} \emph{(§9.9)} & insert / mark
joined with \passthrough{\lstinline!checkpointHash!} in
\passthrough{\lstinline!children!} \\
\passthrough{\lstinline!run.end!} & set
\passthrough{\lstinline!result!}; any subsequent non-meta entry = fold
error \\
\end{longtable}

The \passthrough{\lstinline!manifest.ack!},
\passthrough{\lstinline!child.spawn!}, and
\passthrough{\lstinline!child.join!} entries are in the §4.3 schema
because this appendix requires them: they existed implicitly in §9.9 and
§10.3, and the fold's totality rule forces them explicit -- a small
example of the appendix doing its job.

\hypertarget{commitments-by-reference}{%
\subsection{Commitments by reference}\label{commitments-by-reference}}

Anything with its own authenticated structure enters the transcript as a
\textbf{root, not a body}. Hook views commit as the Autobase view core's
\passthrough{\lstinline!(key, length, viewRoot)!}; the transcript does
not re-serialize the linearized view, because Autobase's own
deterministic apply and checkpointing already authenticate it, and
re-encoding an authenticated structure is a second chance to disagree.
Children commit by their checkpoint hashes (§7.2), manifests by their
amendment-chain head, payloads by the
\passthrough{\lstinline!PayloadRef!} hashes already in the journal. The
transcript is thus a commitment \emph{of commitments}, small, and with
every leaf independently verifiable through the structure that owns it.

\hypertarget{the-canonical-encoding-profile}{%
\subsection{The canonical encoding
profile}\label{the-canonical-encoding-profile}}

The reference codec is \passthrough{\lstinline!compact-encoding!}, the
Pear ecosystem's deterministic binary codec, chosen so a conformant
implementation exists wherever the substrate does, restricted to a
profile that removes every degree of representational freedom:

\begin{enumerate}
\def\labelenumi{\arabic{enumi}.}
\tightlist
\item
  \textbf{No floats, anywhere.} HLCs encode as
  \passthrough{\lstinline!(u64 physical, u32 logical)!}; all counters as
  unsigned varints; amounts and durations as integers in declared base
  units.
\item
  \textbf{Fixed field order} as declared in A.2, all fields always
  present; \passthrough{\lstinline!null!} is an explicit one-byte tag,
  never an omitted field; optionality is representable in exactly one
  way.
\item
  \textbf{Sets encode sorted.} \passthrough{\lstinline!pending!},
  \passthrough{\lstinline!sleeps!} by \passthrough{\lstinline!seq!};
  \passthrough{\lstinline!hooks!} by bytewise
  \passthrough{\lstinline!hookId!}; \passthrough{\lstinline!children!}
  by bytewise \passthrough{\lstinline!coreKey!}. Duplicate sort keys are
  a fold error (A.3), so ordering is total.
\item
  \textbf{Byte strings are bytes.}
  \passthrough{\lstinline!stepId!}/\passthrough{\lstinline!hookId!} are
  length-prefixed byte strings compared and encoded bytewise; no Unicode
  normalization is applied or permitted; identity is octets, and any
  human-legibility convention lives above this layer.
\item
  \textbf{No indefinite lengths, no padding, no alignment.} Every
  container is length-prefixed with the minimal varint.
\item
  \textbf{One profile per \passthrough{\lstinline!wc!} version.} The
  profile is closed: an encoder emitting anything outside it is
  nonconformant even if a decoder would accept it.
\end{enumerate}

Rules 1--5 make \passthrough{\lstinline!canonicalEncode!} injective on
\passthrough{\lstinline!TranscriptState!} up to value equality; rule 6
makes ``same state, different bytes'' a versioning event rather than an
interoperability surprise.

\hypertarget{hash-construction-and-why-excluding-frames-is-sound}{%
\subsection{Hash construction and why excluding frames is
sound}\label{hash-construction-and-why-excluding-frames-is-sound}}

\begin{align*}
\mathit{transcriptHash} = \mathrm{blake3.deriveKey}(\text{"p2p-world transcript v0 "} + wc,\ \mathrm{canonicalEncode}(\mathit{state}))
\end{align*}

Domain separation via BLAKE3's derive-key mode (distinct context strings
for transcript, step chain, custody bindings §12.1, idempotency keys
§8.6) ensures no artifact of one subsystem can be replayed as another's.

The soundness argument for excluding suspended frames is the determinism
contract, and it deserves to be stated with its conditions showing
rather than as an article of faith:

\begin{quote}
\textbf{Proposition (conditional).} For a fixed bundle, a fixed
transcript of settled outcomes, and an execution environment conforming
to the determinism profile below, the orchestrator's suspended
continuation is uniquely determined.
\end{quote}

The unconditional version (``deterministic replay just works'') is what
durable-execution marketing says; the conditional version is what
engineering can defend, because JavaScript environments leak ambient
nondeterminism through channels no directive intercepts. The
\textbf{determinism profile} is therefore a normative part of
\passthrough{\lstinline!worldContract!}, with three obligations:

\textbf{D1: The journal is the scheduler.} When an orchestrator awaits
multiple pending steps or hooks concurrently
(\passthrough{\lstinline!Promise.all!},
\passthrough{\lstinline!Promise.race!}), original execution settles
races by wall-clock arrival, but replay must not. Conformant engines
deliver step results and hook resolutions to the orchestrator \emph{in
journal sequence order}, and the race's winner is thereby a journal
fact, not a scheduler fact. This single rule eliminates the largest
real-world divergence class (microtask interleaving) and is the reason
\passthrough{\lstinline!step.end!} entries carry their own
\passthrough{\lstinline!seq!} rather than trusting arrival order.

\textbf{D2: The ambient surface is closed.} The bundler pass that
already rewrites \passthrough{\lstinline!Date.now()!} and
\passthrough{\lstinline!Math.random()!} onto the tape extends to every
remaining implementation-defined surface: transcendental
\passthrough{\lstinline!Math.*!} functions (whose precision the language
spec leaves to the engine) are rewritten to a deterministic software
implementation shipped \emph{in the bundle}; locale-sensitive APIs
(\passthrough{\lstinline!Intl!},
\passthrough{\lstinline!toLocaleString!}) and other environment probes
are either rejected at build time or tape-recorded;
engine-version-sensitive tables (Unicode property escapes in regex) are
pinned by the profile version. The principle: an orchestrator conforms
not because its author was careful but because the build refuses the
surfaces where care is required.

\textbf{D3: Everything else is tape} (A.9).

\begin{longtable}[]{@{}
  >{\raggedright\arraybackslash}p{(\columnwidth - 4\tabcolsep) * \real{0.3333}}
  >{\raggedright\arraybackslash}p{(\columnwidth - 4\tabcolsep) * \real{0.3333}}
  >{\raggedright\arraybackslash}p{(\columnwidth - 4\tabcolsep) * \real{0.3333}}@{}}
\caption{Determinism-profile escape classes and their obligations
(A.6).}\tabularnewline
\toprule\noalign{}
\begin{minipage}[b]{\linewidth}\raggedright
Escape class
\end{minipage} & \begin{minipage}[b]{\linewidth}\raggedright
Example
\end{minipage} & \begin{minipage}[b]{\linewidth}\raggedright
Obligation
\end{minipage} \\
\midrule\noalign{}
\endfirsthead
\toprule\noalign{}
\begin{minipage}[b]{\linewidth}\raggedright
Escape class
\end{minipage} & \begin{minipage}[b]{\linewidth}\raggedright
Example
\end{minipage} & \begin{minipage}[b]{\linewidth}\raggedright
Obligation
\end{minipage} \\
\midrule\noalign{}
\endhead
\bottomrule\noalign{}
\endlastfoot
Concurrency ordering & \passthrough{\lstinline!Promise.race!} over two
pending steps & D1: delivery in journal order \\
Approximated math & \passthrough{\lstinline!Math.sin!} differing in last
ulp across engines & D2, deterministic softfloat in bundle \\
Environment probes & \passthrough{\lstinline!Intl.DateTimeFormat!},
locale strings & D2: build-time rejection or tape \\
Engine data tables & regex Unicode classes across versions & D2: pinned
by profile version \\
Declared nondeterminism & time, randomness, generated IDs & D3: the
tape \\
\end{longtable}

Under the profile, frames are a \emph{derived} quantity; committing to a
derived quantity's engine-specific representation would subtract
engine-independence while adding no binding power, so the transcript
commits to the determinism function's inputs and the attestation check
audits its outputs. And the design degrades safely on the profile's own
failure: a leak the profile missed does not corrupt state; it produces a
\passthrough{\lstinline!TranscriptDivergence!} at the next verification
seam (fail-detected, never fail-silent) at which point \textbf{engine
pinning returns as the recovery attempt}: the resumer re-executes under
the exact engine build named in the last commitment, which reproduces
the committed transcript if the divergence was engine skew, and fails
detected again if the leak was ambient -- the case §7.1 names as the
honest floor. Engine pinning is therefore not a correctness
mechanism in the ordinary path (§9.5): it is the designated recovery
attempt whenever the cheaper guarantee is caught leaking -- an attempt,
per §7.1, not a guarantee -- and every such divergence is a minimizable
bug report against the profile.

\hypertarget{the-next-intent-record-where-the-engine-enters}{%
\subsection{The next-intent record: where the engine
enters}\label{the-next-intent-record-where-the-engine-enters}}

Everything above is engine-independent by design, which is precisely why
it cannot verify an engine (§7). The next-intent record is the
deliberate exception. It is \textbf{not} part of
\passthrough{\lstinline!TranscriptState!} and never enters the fold; it
is \emph{engine output}, computed from the live continuation at a
suspension boundary, and carried in the
\passthrough{\lstinline!transcript.commit!} entry (itself meta-excluded
from the fold, so no circularity arises). Its canonical form:
\passthrough{\lstinline!\{ kind, callSite, callIndex, argsHash \}!},
encoded under the A.5 profile, where \passthrough{\lstinline!callSite!}
is the bundler-assigned stable identifier of the awaited call
expression, \passthrough{\lstinline!callIndex!} is the ordinal of this
call in the orchestrator's deterministic call sequence (making loops and
re-entries unambiguous), and \passthrough{\lstinline!argsHash!} digests
the canonically encoded arguments. Conformance for intent derivation is
behavioral like everything else in this appendix: given a journal prefix
and bundle, a conformant engine's replay must produce the recorded next
intent, and A.11's vectors include prefixes whose correct outcome is a
specified \passthrough{\lstinline!TranscriptDivergence!}, because
agreeing on \emph{rejection} is as load-bearing as agreeing on intent.

\hypertarget{the-replay-capsule-what-resumption-actually-reads}{%
\subsection{The replay capsule: what resumption actually
reads}\label{the-replay-capsule-what-resumption-actually-reads}}

\passthrough{\lstinline!TranscriptState!} proves; the capsule supplies
(§7.2). Its canonical form \textbf{{[}P{]}}: a length-prefixed sequence,
in canonical stream order, of settled-operation records
\passthrough{\lstinline!(seq, kind, callSite, callIndex, argsHash, outcome)!}
(\passthrough{\lstinline!outcome!} inline under the A.5 encoding when
small, a \passthrough{\lstinline!PayloadRef!} otherwise) followed by the
resolved-hook table
\passthrough{\lstinline!(hookId, outcome, baseCommitment)!}, the
consumed prefix of the entropy/time tape (values, not just the cursor),
and the manifest amendment chain up to the checkpoint. Everything replay
consumes and nothing it does not: signatures, Merkle nodes, and wire
framing are stripped, which is the compaction. Two obligations attach,
and one tempting claim must be rejected first: ``the capsule's fold
reproduces the snapshot'' is false: the capsule omits absorbed attempts
(which live in the snapshot's \passthrough{\lstinline!stepChain!}) and
all active state, so no fold of capsule records alone can reproduce
\passthrough{\lstinline!TranscriptState!}. The precise relationship
\textbf{{[}P{]}}: \emph{Integrity} is by co-signed hash: the checkpoint
quorum independently signs both object hashes, so a failover replica
verifies each object against the checkpoint it already trusts, needing
no archival peer; the snapshot supplies accumulated commitments and
active state, the capsule supplies replay values, and cross-consistency
between them (capsule outcome commitments agreeing with the snapshot's
accumulator) is not resume-time checkable without the full prefix; it is
validated by archival replay in the conformance corpus (A.11), which is
where a lying checkpoint quorum would be caught. \emph{Sufficiency:} a
conformant engine given snapshot + capsule + suffix must reach the same
next intent as one given the full journal; A.11's vectors include
capsule-bootstrapped variants of every resume case for this reason.
\textbf{The executable mapping {[}P{]}.} The property list above is a
specification assertion until the mapping is code-shaped, so the
\textbf{canonical constructor} is given as a deterministic single pass,

\begin{lstlisting}
function buildCapsule(stream: Entry[], p: number, wc: WorldContract): Capsule {
  const cap = new Capsule(wc);
  for (const e of stream.slice(0, p)) switch (e.t) {
    case "step.end":
      cap.settle(e.seq, intentTupleOf(begin(e)), outcomeOf(e)); break;
    case "step.error":       // absorbed retries: final attempt only;
      if (!e.willRetry)      // earlier attempts live in the SNAPSHOT's
        cap.settle(e.seq, intentTupleOf(begin(e)), errorOf(e));
      break;                 // stepChain accumulator, not here
    case "hook.receive":
      cap.resolveHook(e.hookId, outcomeOf(e), baseOf(awaitOf(e))); break;
    case "child.join":       // child's own capsule fetched by this hash
      cap.joinChild(e.coreKey, e.checkpointHash); break;
    case "manifest.ack":
      cap.amendManifest(e.chainHead); break;
    case "epoch.open":       // synthetic entries: the closure chain IS
      cap.closeEpoch(e); break;  // capsule content -- resume cannot
                             // verify chain linkage without it
    case "tape.segment":
      cap.appendTape(e.values); break;
  }
  return cap;                // open operations are NOT here: they are
}                            // snapshot state, replayed live (S7.2)
\end{lstlisting}

and the \textbf{consumer interface} is the seam that makes ``snapshot +
capsule + suffix behaves like the full journal'' testable rather than
aspirational: replay never touches entries directly but reads a
\passthrough{\lstinline!ReplaySource!},
\passthrough{\lstinline!\{ outcomeAt(callIndex), hookOutcome(hookId), childCheckpoint(coreKey), tapeAt(cursor), manifestAt(seq), closureAt(epoch) \}!},
with two conforming implementations,
\passthrough{\lstinline!JournalSource(entries)!} and
\passthrough{\lstinline!CapsuleSource(snapshot, capsule, suffix)!} -- the
snapshot initializing the pending set, open sleeps and hooks, live
children, epoch position, \passthrough{\lstinline!stepChain!}, and
entropy cursor; the capsule supplying settled values; the suffix
continuing from the checkpoint. Engine conformance is defined over the
interface: identical behavior across sources, which is precisely what
A.11's capsule-bootstrapped vectors exercise. Cadence composes
trivially: a snapshot and capsule at checkpoint position
\passthrough{\lstinline!p!} plus the suffix
\passthrough{\lstinline![p, head)!} are a complete source.
\textbf{Encryption:} the capsule is itself an enveloped object on the
distribution plane (content-addressed ciphertext under a capsule DEK
wrapped to the run KEK) and outcome
\passthrough{\lstinline!PayloadRef!}s inside it stay enveloped just as
in the journal; consuming a capsule therefore requires Ring-2 access by
construction, which is consistent rather than incidental:
\emph{failover-capable} has always implied key possession, and the
capsule adds no new plaintext surface that the journal did not already
have.

The capsule's size is O(settled operations); its
\passthrough{\lstinline!PayloadRef!}s inherit the retention policy of
the payloads they name -- a capsule does not resurrect what erasure (§9.3)
removed -- and a run relying on redacted replay must satisfy §9.3's
control-flow condition regardless of bootstrap path.

\hypertarget{the-entropy-and-time-tape}{%
\subsection{The entropy and time tape}\label{the-entropy-and-time-tape}}

All nondeterminism the orchestrator consumes (intercepted
\passthrough{\lstinline!Date.now()!},
\passthrough{\lstinline!Math.random()!}, generated IDs) is a
\textbf{tape of journal facts}: each value is recorded at first
execution (inline in the consuming entry or in a dedicated tape segment
for high-volume consumers), and replay reads the tape rather than the
world. The transcript commits only to
\passthrough{\lstinline!entropyCursor!}, the count consumed, because the
values themselves are already committed by the journal blocks that carry
them. Consequently sleep expiry, challenge windows, and every other
temporal judgment inside the fold compares \emph{recorded HLCs to
recorded HLCs}; the fold never reads a clock, which is what makes it
replayable identically in ten years.

\hypertarget{versioning-and-bridge-attestations}{%
\subsection{Versioning and bridge
attestations}\label{versioning-and-bridge-attestations}}

\passthrough{\lstinline!wc!} names this entire appendix: state shape,
transition table, encoding profile, hash domains. A running workflow
migrates versions by an explicit \textbf{bridge attestation}: at a
suspension boundary the executor appends a
\passthrough{\lstinline!transcript.commit!} carrying \emph{two} hashes
of the same \passthrough{\lstinline!seq!} (the state folded and encoded
under \passthrough{\lstinline!wc!} and under
\passthrough{\lstinline!wc'!}) after which the run proceeds under
\passthrough{\lstinline!wc'!}. Verifiers spanning the bridge check both;
history below the bridge remains verifiable under the version that wrote
it. Version skew thereby becomes a signed, located, co-verifiable
journal event -- the same treatment §8 gave executor changes, applied to
the specification itself.

\hypertarget{conformance}{%
\subsection{Conformance}\label{conformance}}

Conformance is behavioral, not asserted: an implementation is conformant
iff it reproduces the published \textbf{test-vector corpus}: journal
prefixes paired with expected \passthrough{\lstinline!TranscriptState!}
encodings and hashes, including adversarial vectors (duplicate pending
keys, unknown entry types, boundary HLCs, empty and maximal sets) that
must produce the \emph{specified fold errors}, since agreeing on failure
matters as much as agreeing on success. Development runs a
\textbf{differential harness}: two engines fold randomized journals and
diff hashes, with any divergence minimized to a reportable vector. And
production needs no separate monitor, because §7.1's resume-time check
makes every handoff between heterogeneous engines a conformance test
executed on real workloads; the specification is audited for free,
forever, by the protocol that depends on it.

\begin{center}\rule{0.5\linewidth}{0.5pt}\end{center}

\end{document}